\documentclass[acmsmall,screen]{acmart}

\AtBeginDocument{%
  }

\setcopyright{cc}
\setcctype{by}
\acmJournal{PACMSE}
\acmYear{2026}
\acmVolume{3}
\acmNumber{FSE}
\acmArticle{FSE012}
\acmMonth{7}
\acmDOI{10.1145/3797149}
\acmSubmissionID{fse26maina-p227-p}
\received{2025-09-05}
\received[accepted]{2025-12-22}

\usepackage[english]{babel} 
\usepackage{cleveref}
\crefname{section}{}{\S\S}
\crefdefaultlabelformat{\S#2#1#3}
\usepackage{xspace}
\usepackage{comment}
\usepackage[normalem]{ulem}
\usepackage{enumitem}
\setlist{nolistsep}
\usepackage{listings}
\usepackage{color}
\usepackage[font=footnotesize]{caption}
\graphicspath{{figures/}}

\usepackage[linesnumbered,ruled,vlined]{algorithm2e}
\usepackage{xcolor}
\SetAlgoNlRelativeSize{0} %
\SetAlgoNlRelativeSize{-1} %
\SetKwComment{Comment}{}{} %

\usepackage{amsmath,pifont}
\usepackage{multirow}
\usepackage{textcomp}
\usepackage[most]{tcolorbox}
\usepackage{hyperref}
\usepackage{pifont}   

\newcommand{\cmark}{\ding{51}}%
\newcommand{\xmark}{\ding{55}}%
\newcommand{\yx}[1]{}
\newcommand{\hxq}[1]{}

\newcommand{\SYS}{TSGuard\xspace}
\begin{document}

\title[\SYS]{\SYS: Automated User-Centric Incident Diagnosis for AI Workloads in the Cloud}

\author{Yitao Yang}
\authornote{Part of this work was done during an internship at Microsoft Research.}
\orcid{0009-0000-1332-7342}
\affiliation{%
  \institution{The Chinese University of Hong Kong}
  \city{Hong Kong}
  \country{Hong Kong}
}
\email{ytyang@cse.cuhk.edu.hk}

\author{Yangtao Deng}
\orcid{0000-0003-4072-7696}
\affiliation{%
  \institution{The Chinese University of Hong Kong}
  \city{Hong Kong}
  \country{Hong Kong}
}
\email{ytdeng25@cse.cuhk.edu.hk}

\author{Yifan Xiong}
\orcid{0000-0001-9056-1386}
\affiliation{%
  \institution{Microsoft Research}
  \city{Vancouver}
  \country{Canada}
}
\email{yifan.xiong@microsoft.com}

\author{Baochun Li}
\orcid{0000-0003-2404-0974}
\affiliation{%
  \institution{University of Toronto}
  \city{Toronto}
  \country{Canada}
}
\email{bli@ece.toronto.edu}

\author{Hong Xu}
\orcid{0000-0002-9359-9571}
\affiliation{%
  \institution{The Chinese University of Hong Kong}
  \city{Hong Kong}
  \country{Hong Kong}
}
\email{hongxu@cuhk.edu.hk}

\author{Peng Cheng}
\orcid{0000-0003-4014-4757}
\affiliation{%
  \institution{Microsoft Research}
  \city{Redmond}
  \country{USA}
}
\email{pengc@microsoft.com}

\renewcommand{\shortauthors}{Yang et al.}

\begin{abstract}
AI workloads incur frequent failures and incidents from the underlying infrastructure. 
The current incident management workflow follows a provider-centric paradigm, where users report incidents to the infrastructure provider who then conducts troubleshooting.
Due to the large number of incidents and the manual nature of the troubleshooting process, the provider often takes several days to resolve an incident, resulting in operational delays and productivity loss.

To address these challenges, we present TSGuard, a user-centric multi-agent system that delivers immediate incident diagnosis to users who deploy the workloads.
The core innovation of TSGuard is twofold: (1) constructing domain-specific knowledge bases by mining historical on-call experiences in the offline phase, and (2) mimicking human expert diagnosis via structured reasoning and iterative trial-and-error in the online phase.
Evaluation using production incident records from Microsoft Azure demonstrates that TSGuard significantly outperforms state-of-the-art baselines, improving diagnostic accuracy by 19.8\%. Furthermore, TSGuard reduces the average verification time by 63.4\% compared to the sequential execution baseline.
\end{abstract}




\maketitle

\section{Introduction}
\label{sec:introduction}


With the exponential growth of AI workloads~\cite{OpenAIGPT-4o, dubey2024llama, achiam2023gpt,yang2024qwen2}, operational failures~\cite{cui2025xputimer,dong2025evolution} have become increasingly prevalent in AI infrastructure in the cloud.
Moreover, due to the synchronous nature of these workloads, failures in large-scale AI infrastructure can have a disproportionately larger impact~\cite{deng2025minder, wang2023gemini}.
Such failures, if not handled promptly, can lead to significant disruptions, causing substantial waste of computational and financial resources of users.
For instance, Meta reported 466 job interruptions during a 54-day Llama3.1 405B pre-training session~\cite{dubey2024llama}, which wasted over 2.12 million H100 GPU hours with an estimated cost of over 18 million dollars based on Azure cloud GPU pricing~\cite{azure_cloud_gpu_service}.
%

\begin{figure}[t]
    \centering
    \includegraphics[width=0.8\linewidth]{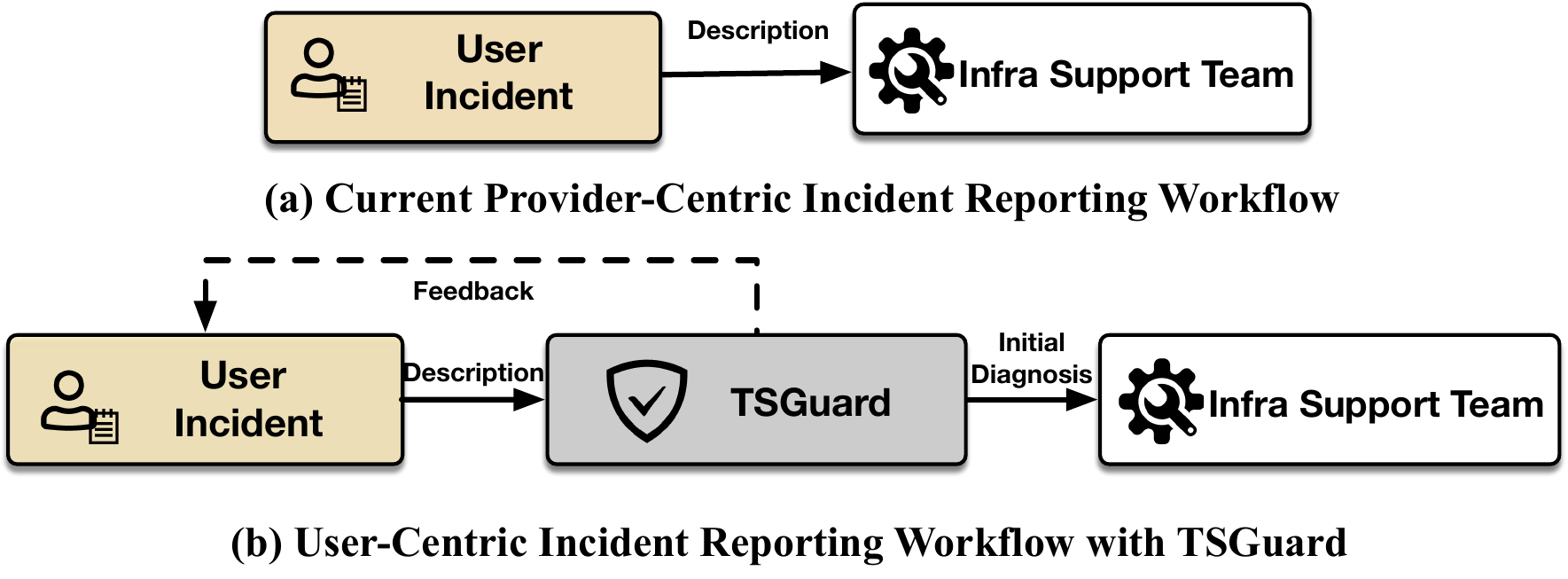}
    \vspace{-1em}
    \caption{Comparison of provider-centric and user-centric incident reporting workflow.}
    \label{fig:workflow}
    \vspace{-2em}
\end{figure}

We focus on the cloud scenario where the infrastructure is provided to users for deploying their AI workloads. 
Note that users here can be both internal first-party teams that deploy the provider's own AI workloads, and external third-party users who rent the cloud.
They lack direct control of the hardware and comprehensive knowledge about the software stack and the whole infrastructure, while providers are restricted from seeing the specific task settings. 
These constraints shape the current incident management workflow, illustrated in Figure~\ref{fig:workflow}(a), which is largely \textit{provider-centric}: users report incidents often with very limited client-side diagnostic information to the provider, waiting for the provider to identify and resolve the issue.
%

This workflow, however, suffers from significant limitations due to an inherent knowledge gap between users and the provider, leading to low efficiency and inadequate diagnostic feedback. 
Users typically lack the technical expertise to provide relevant and detailed diagnostic information, resulting in incomplete and ambiguous incident reports. Further, the effectiveness of this workflow depends heavily on subsequent communication between the user and the on-call engineers (OCEs) handling the case, leading to additional delays.
Our one-year analysis of the production incidents in Microsoft Azure underscores this inefficiency: the median time to mitigate (TTM) was 52.5 hours, with a mean TTM of 83.0 hours.

Recent advancements in natural language understanding and tool utilization capabilities~\cite{schick2023toolformer,yang2024gpt4tools} of large language models (LLMs) show great potential in bridging this user-provider gap to enable more efficient and timely diagnosis. 
Several prior works have made initial explorations for using LLMs in cloud incidents, with a primary focus on provider-side automation.
Their designs essentially map the incident description to the potential root cause using an LLM, providing a one-shot prediction without any intermediate reasoning or feedback~\cite{ahmed2023recommending,RCACopilot,han2024potential, zhang2024automated,an2024nissist,jiang2024xpert,zhang2023pace,xie2024cloud}.

Our analysis reveals that these provider-centric schemes overlook the user's potential in incident management.
Empowering users to self-diagnose incidents first and contact the provider only when needed could significantly reduce ticket volumes and operational labor, leading to a more efficient incident management process. 
Even for provider-related issues, the user's initial investigation can provide more precise incident descriptions, which can facilitate faster resolution.
However, due to users' varying levels of technical expertise and skills, current provider-centric workflow is unable to leverage this potential and places all the burden on the provider.
%
Second, existing methods also oversimplify the unique challenges of AI workloads and infrastructures. 
We find that the root cause distribution in AI workloads is heavily skewed toward hardware failures, particularly GPU-related issues, which constitute over 50\% of incidents and exhibit high recurrence rates.
In contrast, over 60\% of failures in traditional cloud workloads stem from code and dependency issues~\cite{ghosh2022fight}, highlighting a fundamentally distinct failure pattern. 
These hardware failures necessitate physical interaction and metrics for diagnosis~\cite{deng2025minder,DCGM,xiong2024superbench}, which are not used or supported by existing provider-centric diagnostic systems~\cite{RCACopilot,ahmed2023recommending,huang2024faultprofit}.


In this paper, we propose \SYS, a user-centric system that automates AI incident diagnosis and reporting (Figure~\ref{fig:workflow}(b)). 
\SYS provides two major benefits: (1) for users, it delivers real-time diagnostic feedback; (2) for providers, it automates initial diagnosis on the user side and generates preliminary diagnostic reports for unresolved cases. 
These capabilities enable \SYS to reduce the burden on OCEs and expedite incident resolution.

Achieving these goals is challenging, primarily for two reasons:
(1) off-the-shelf LLMs lack domain-specific knowledge of internal infrastructure for accurate incident diagnosis;
(2) diagnosing AI incidents solely based on symptoms is insufficient, as root cause confirmation requires verification evidence~\cite{xiong2024superbench}.
\SYS tackles these challenges in two complementary phases.
During the offline phase, \SYS constructs three structured knowledge components from historical incident records.
In online diagnosis, \SYS employs a tiered diagnostic pipeline to automatically identify potential root causes through systematic hypothesis generation and verification, delivering real-time diagnostic feedback to users.
For unresolved incidents, \SYS escalates them with preliminary diagnostic reports to the infrastructure support team, thereby reducing their burden.

We implement a prototype of \SYS and evaluate its performance using one-year production incident records from Microsoft Azure. The evaluation demonstrates that \SYS achieves average Micro F1 and Macro F1 scores of 0.854 and 0.816, respectively. These scores significantly outperform those of the state-of-the-art (SOTA) baseline, RCACopilot~\cite{RCACopilot}, by 19.8\% and 43.6\%.
Moreover, \SYS reduces the average verification time by 63.4\% compared to the sequential benchmark execution baseline.

Our contributions are summarized as follows:

\noindent $\bullet$ We identify inefficiencies in the current provider-centric incident management workflows for AI workloads, where inherent knowledge gaps between users and infrastructure impact diagnostic accuracy and resolution times.

\noindent $\bullet$ We analyze existing LLM-based incident diagnosis systems and expose their limitations in addressing the complexities of AI workload incidents.

\noindent $\bullet$ We propose \SYS, a user-centric system for automated AI workload incident diagnosis.
\SYS leverages domain-specific knowledge to mimic the reasoning and diagnostic process of OCEs and streamline incident diagnosis and reporting.

\noindent $\bullet$ We implement \SYS prototype and evaluate its performance using real-world incident records, demonstrating its effectiveness compared to various baselines.

\section{Background and Motivation}
\label{sec:motivation}

We start by introducing our production incident analysis of AI infrastructure and the limitations of the current provider-centric workflow. We then discuss the opportunities and challenges for leveraging LLMs in incident diagnosis.

\subsection{AI Workload Incidents in the Wild}
\label{sec:ai_infra_incident}
While prior studies have extensively investigated incidents in traditional cloud workloads~\cite{RCACopilot, dogga2023autoarts, ahmed2023recommending, zhang2024automated, jin2023assess, ghosh2022fight}, AI workload incidents present substantially more complex challenges. 
To establish this complexity, we collect and analyze $\sim$1,300 real-world incident data from production GPU clusters serving users' AI workloads at Microsoft Azure spanning a one-year period (2023-04 to 2024-03).
The incidents are either reported by users or detected by the cloud provider's monitoring systems, and contain error symptoms (error messages, logs, etc.), root causes, postmortem discussions of OCEs, and resolution steps. 
More details about data collection and screening are provided in \cref{sec:data_collection}.

{\bf Root Cause Characterization.}
Modern AI infrastructure is built upon advanced hardware like high-performance GPUs (e.g., NVIDIA H100~\cite{nvidia_h100}, AMD MI200~\cite{amd_mi200}) and specialized interconnects (NVLink~\cite{nvidia_nvlink}, InfiniBand~\cite{nvidia_quantum_IB}). Unlike traditional infrastructure 
 that primarily rely on relatively stable CPU, memory, and storage resources, AI infrastructure faces more demanding conditions, especially during large-scale, extended model training with thousands of GPUs~\cite{dubey2024llama}.

\begin{figure}[t]
    \centering
    \begin{minipage}{.46\linewidth}  %
      \centering
      \includegraphics[width=1\linewidth]{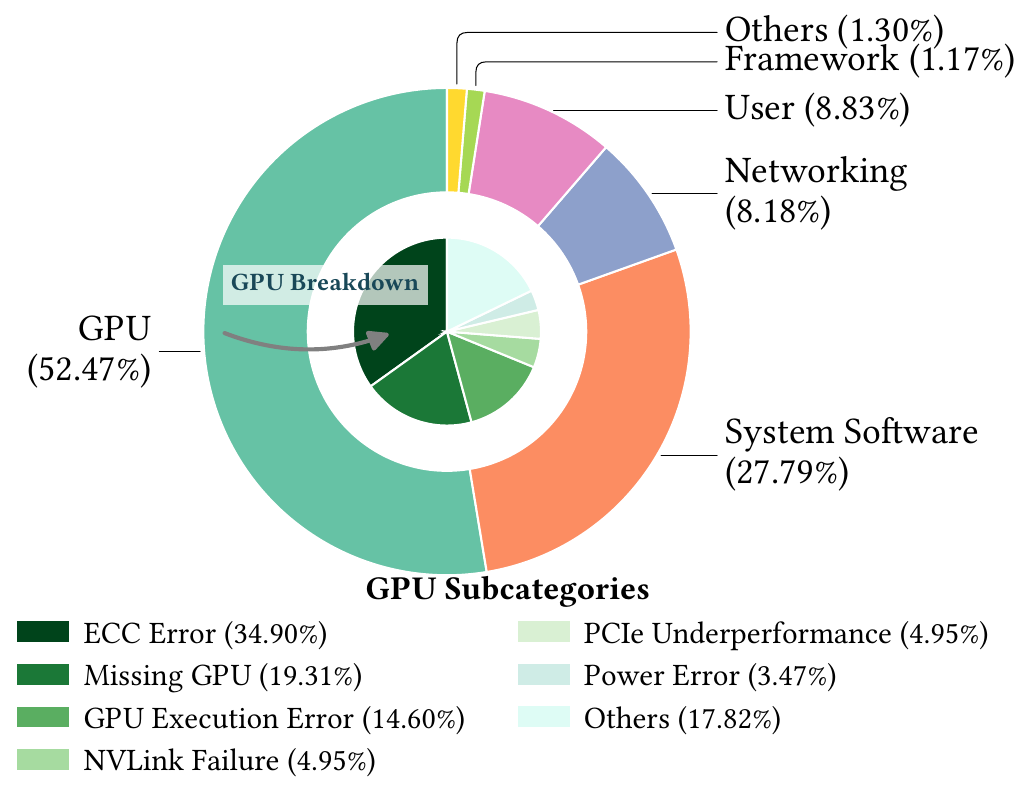}
        \vspace{-2.5em}
      \caption{Breakdown of root cause categories for AI infrastructure in Microsoft Azure. Detailed definitions of each category are provided in \cref{sec:taxonomy}.} %
      \label{fig:category_breakdown}
    \end{minipage}%
    \hspace{0.15cm}
    \begin{minipage}{.34\linewidth}  %
        \centering
        \includegraphics[width=\linewidth]{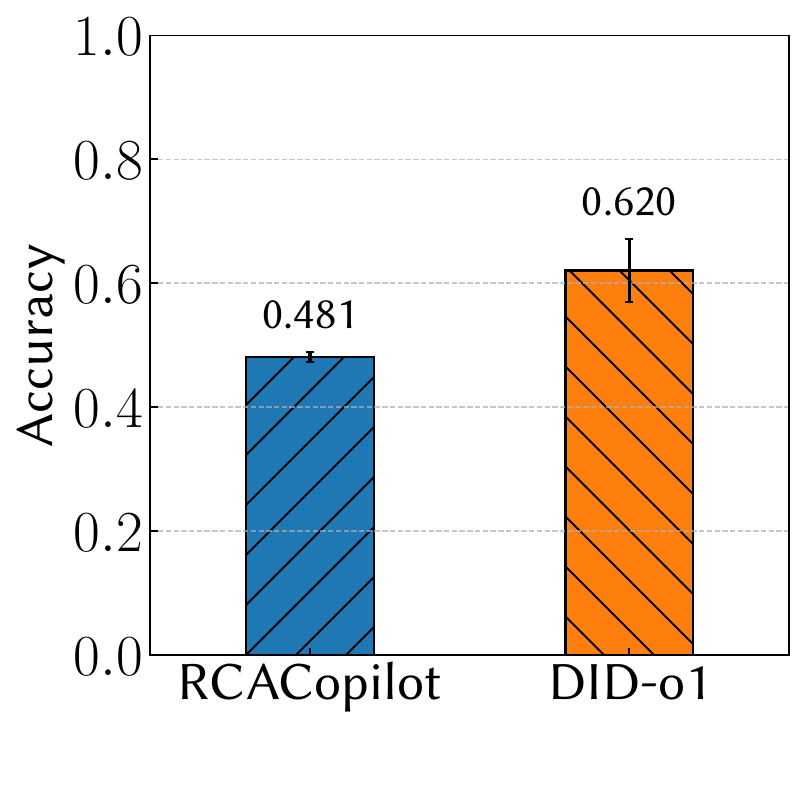}
        \vspace{-2.5em}
        \caption{Performance of two LLM-based diagnosis systems on AI workload incidents.}
        \label{fig:motivation_accuracy}
    \end{minipage}
    \vspace{-1.3em}

\end{figure}

As a result, AI workload incidents exhibit distinctive root cause distributions compared to cloud workloads.
Our one-year analysis (Figure~\ref{fig:category_breakdown}) shows GPU-related failures comprise 52.47\% of AI incidents. Among them, ECC errors (34.90\%), missing GPUs (19.31\%), and execution errors (14.60\%) are the primary contributors.
Conversely, production cloud workloads primarily face code/config bugs ($\sim$40\%), dependency failures (16.4\%), and general infrastructure issues (15.6\%) according to~\cite{ghosh2022fight}.
This disparity highlights the distinct challenges of managing AI workload incidents, as the related GPU, IB, and even software issues (CUDA, PyTorch, etc.) are undergoing rapid development and require much expertise to understand~\cite{xiong2024superbench}.

{\bf Failure Pattern Analysis.}
We analyze the recurrence rate of incidents over the year to identify persistent failure patterns in AI workloads (Table~\ref{tab:incident_repetition}).
The recurrence rate, calculated as total incidents in a category within the year divided by its distinct fault types, measures the average reoccurrence of each distinct fault type. Higher values indicate more frequent repetition of the same underlying faults.
Our analysis reveals that GPU-related incidents exhibit the highest recurrence rate (8.78), followed by networking (3.15) and system software (2.34).
This pattern stems from the large-scale deployment of GPU clusters with thousands of interconnected devices, where hardware failures and network issues tend to recur systematically. 
In contrast, conventional workload incidents are typically caused by code or software errors. These errors often recur within a short time frame but tend to stop once the software is patched or updated, as noted in~\cite{RCACopilot}.

{\bf Non-Indicative Symptoms and Unreliable Symptomatic Diagnosis.}
Unlike traditional systems, the symptoms of AI incidents are often weak indicators of root causes. In traditional infrastructure, a symptom can map cleanly to a specific issue (e.g., a DNS resolution failure in a mail server typically signals UDP hub port exhaustion~\cite{RCACopilot}). By contrast, AI workloads exhibit many-to-many symptom–cause relations: a single fault can produce multiple, seemingly unrelated signals (e.g., reduced CPU utilization and NVLink bandwidth with error codes), while a single symptom (e.g., a generic GPU error) may arise from hardware, driver, or resource issues~\cite{deng2025minder}. This necessitates diagnosis beyond one-shot symptom matching.

For example, a user reported a CUDA error ``\texttt{invalid device ordinal}'' with an NVIDIA \texttt{Xid 119} in \texttt{kern.log}, alongside drops in CPU utilization and NIC throughput. Rebooting temporarily cleared the error, but it recurred after training resumed. The final root cause was not GPU hardware failure, but a CUDA driver version mismatch between the Docker image and the VM.

\renewcommand{\arraystretch}{1} %
\setlength{\tabcolsep}{2pt}       %
\begin{figure}[tbp]
    \centering
    \begin{minipage}{.39\linewidth}  %
    \centering
    \small
    \captionof{table}{Per-category recurrent failure frequency (higher values indicate more frequent occurrences).}
    \label{tab:incident_repetition}
    \vspace{-1em}
    \resizebox{0.75\columnwidth}{!}{%
    \begin{tabular}{cc}
        \hline
        \begin{tabular}[c]{@{}c@{}}Main \\ Category\end{tabular} & \begin{tabular}[c]{@{}c@{}}Recurrence \\ Rate\end{tabular} \\ \hline
 GPU                                                               & 8.78                                                                \\
 System Software                                                   & 2.34                                                                \\
 Networking                                                        & 3.15                                                                \\
 User Apps                                                         & 1.86                                                                \\
 Framework                                                         & 1.12                                                                \\
 Other                                                             & 1.11                                                                \\ \hline
        \end{tabular}
 }
    \end{minipage}%
    \hspace{0.1cm}
    \begin{minipage}{.47\linewidth}  %
    \centering
    \small
    \captionof{table}{Average count of semantically distinct incident descriptions per root cause category.}
    \label{tab:distinct_description}
    \resizebox{0.9\columnwidth}{!}{%
      \begin{tabular}{cc}
        \hline
        Root Cause & Distinct Descriptions \\ 
        Category       & (per 10 incidents)    \\ \hline
 ECC Error                            & 5.6                   \\
 NVLink Failure                       & 5.2                   \\
 IB Networking Error                  & 7.6                   \\
 Illegal Memory Access                & 4.8                   \\ \hline
        \end{tabular}
 }
    \end{minipage}
    
    \vspace{-2em}
\end{figure}

\subsection{Limitations of Current Workflow}
\label{sec:motivation_current_workflow}

{\bf Current Incident Management Workflow.}
In the typical incident lifecycle, users initiate the process by reporting an incident with a description of the issue through a ticketing system. 
The ticket is then routed to the appropriate team for follow-up, i.e., triaging~\cite{gao2020scouts, chen2019continuous, chen2020incidental}. 
OCEs investigate the incident by running additional tests (e.g., NCCL tests~\cite{nccl_tests} and perftests~\cite{perftest}) to reproduce the issue, checking the error logs, and examining the hardware counters for cross-verification.
This process may also include consulting with other teams when needed, as well as collaborating with the user to identify the root cause and implement a resolution~\cite{chen2020incidental,RCACopilot}.

{\bf Limitations.}
The current workflow is \textit{provider-centric}, i.e., designed from the provider's perspective since it is the sole party dealing with the incident here. 
Existing work on automated incident management is also provider-centric~\cite{RCACopilot, dogga2023autoarts, ahmed2023recommending, zhang2024automated, jiang2024xpert, li2023exploring, an2024nissist, zhang2023pace, wang2024rcagent, chen2019continuous, hamadanian2023holistic, shan2024face, wang2024netassistant, xie2024cloud}.
However, this view overlooks the difficulties faced by users, which in turn compromise the overall efficiency of the process. We detail these limitations now.

\textbf{L1: Inefficiency due to Reporting Quality.}
The effectiveness and efficiency of provider-centric workflow heavily depend on the quality of initial incident description reporting and subsequent user communication.
However, we find that the incident description varies significantly among users.
While some provide comprehensive documentation, including detailed error messages, logs, and reproduction steps, others submit minimal information such as basic error stack traces. 
 To quantify the inconsistency in incident descriptions, we randomly selected ten incidents from each specific root cause category and analyzed their descriptions. Using cosine similarity-based clustering~\cite{huang2008similarity}, we calculate the number of unique clusters, representing semantically distinct incident descriptions.
Our analysis (Table~\ref{tab:distinct_description}) reveals that for a single root cause, users provide an average of 4.8 to 7.6 distinct descriptions. This diversity in reporting requires additional communication and troubleshooting, which in turn significantly delays resolution.

\textbf{L2: Hidden Costs due to User Mistakes.}
The current workflow requires the provider to perform full troubleshooting for all reported incidents, regardless of their root cause and origin. 
For incidents caused by user-side misconfigurations or code errors, this leads to much wasted time and resources, as the provider's support efforts are spent on handling issues outside their responsibility. 
This also causes prolonged resolution time for users in turn.

\textbf{L3: Underutilized User Potential in Incident Management.}
If users were able to diagnose the incidents themselves and handle their own mistakes before contacting the provider, they could significantly reduce the time to resolution, and the OCEs' workload could also be reduced. 
Further, even for incidents caused by provider issues, user-side initial investigation can provide valuable information and better incident tickets to the provider, facilitating faster resolution.
Yet, because users possess varying levels of technical expertise and many lack the skills needed to troubleshoot complex issues, the current workflow is unable to leverage this potential and forces all burden on the provider's side. 

\subsection{User-Centric Incident Diagnosis}
\label{sec:challenges}

In light of the above limitations of provider-centric workflow, we propose to develop a \textit{user-centric} incident diagnosis framework that runs on the client side to diagnose incidents as soon as they appear.
It serves the dual benefits of (1) efficiency improvement: identifying user errors without unnecessarily relying on the provider, and (2) diagnosis feedback: providing more comprehensive and useful initial investigations for the users and provider, especially when the incident cannot be resolved by users. Both benefits lead to faster incident resolution for users and reduced workload for the provider compared to the provider-centric workflow.

More specifically, we leverage LLMs with strong natural language understanding and tool usage capabilities~\cite{paranjape2023art,schick2023toolformer,yang2024gpt4tools} to close the user knowledge gap and empower them to diagnose incidents independently like an expert.
The prevalence of text-heavy artifacts in incident workflows (e.g., reports, error logs, traces, and OCE discussions) makes LLMs a natural fit for this diagnostic role.

However, using LLM to solve AI workload incidents still faces key, under-addressed challenges:

{\bf Ch1: Lack of Iterative Feedback and Self-verification Ability.}
Many LLM-based root cause analysis (RCA) systems attempt to map incident descriptions to root causes in a single, non-iterative step using fine-tuning or retrieval augmented generation (RAG)~\cite{RCACopilot, ahmed2023recommending, zhang2024automated, jiang2024xpert, li2023exploring, an2024nissist, zhang2023pace, wang2024rcagent, chen2019continuous, hamadanian2023holistic, shan2024face, wang2024netassistant, xie2024cloud}. This ``one-shot'' paradigm omits the critical processes of hypothesis iteration and evidence-backed confirmation. Consequently, it struggles when initial symptoms are non-indicative, as is common in AI workloads. 
For instance, on our dataset, the direct prediction models RCACopilot (GPT-4o + RAG) and DID-o1 (o1-preview) achieve low accuracies of only 48.1\% and 62.0\%, respectively (Figure~\ref{fig:motivation_accuracy}).
These results reveal a fundamental process limitation: without an embedded mechanism for verification and feedback, diagnoses are less transparent and inherently less trustworthy.

{\bf Ch2: Semantic-Based Retrieval Alone is Unreliable.}
The reliance on semantic retrieval from incident descriptions is a primary source of diagnostic errors. 
Our analysis of RCACopilot's~\cite{RCACopilot} incorrect predictions reveals that its RAG-based design frequently retrieves historical incidents that are semantically similar but contextually and fundamentally irrelevant to the ongoing incident. 
While methods exist to enhance retrieval accuracy~\cite{wang2024searching, sawarkar2024blended}, the fundamental limitations of retrieval-based methods remain since semantic similarity is not a reliable proxy for causal relevance in the AI incident domain (\cref{sec:ai_infra_incident}) and can be attributed to multiple factors~\cite{xiong2024superbench}.
Thus, improving diagnostic coverage and reducing errors remain critical challenges.

{\bf Ch3: Domain-Specific Knowledge is Missing.}
LLMs often lack proprietary, rapidly evolving infrastructure knowledge, which is critical for diagnosing AI incidents with specialized hardware/software.
We showcase such an example here from our empirical dataset. 
In Microsoft Azure, one VM type uses eight InfiniBand (IB) NICs for the data plane and one Ethernet NIC for the control plane.
Due to security limitations, the standard IB tools (i.e., \texttt{ibv\_devinfo}) are unavailable inside containers deployed on these VMs, even under normal operating conditions.
When troubleshooting an incident with \texttt{NCCL WARN NET/IB "Unable to open device mlx5\_1"} error, an LLM (GPT-4o) incorrectly determines hardware failure as the most likely root cause and suggests using \texttt{ibv\_devinfo} to verify the issue. 
The actual root cause, found by OCEs, is a user misconfiguration of the \texttt{NCCL\_IB\_HCA} environment variable, which mistakenly involves the Ethernet device (\texttt{mlx5\_1}). 
The OCEs are aware that \texttt{mlx5\_1} is the designated interface for the control plane, whereas LLMs do not know.

\section{\SYS System}
\label{sec:design}
\begin{figure}[t]
    \centering
    \includegraphics[width=0.5\linewidth]{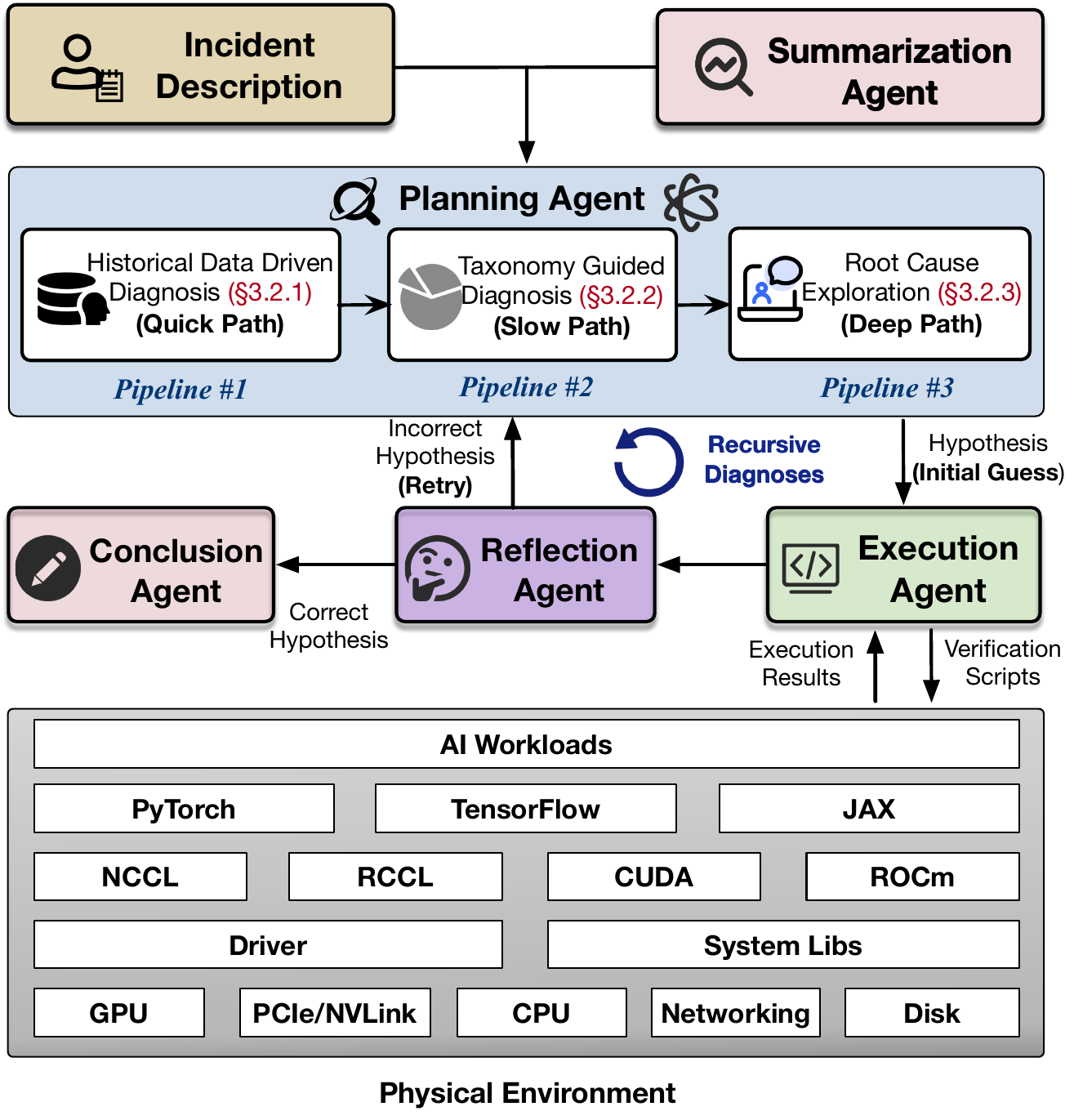}
    \vspace{-3.5mm}
    \caption{Online phase overview in \SYS.}
    \vspace{-5mm}
    \label{fig:Incident Mitigation Pipeline}
\end{figure}

To address the challenges in \cref{sec:motivation}, we propose \SYS, a user-centric multi-agent system for automated incident diagnosis in AI workloads. \SYS features a two-phase architecture: an offline phase for consolidating three types of internal knowledge from historical incidents (\cref{sec:offline_learning}), and an online diagnosis phase that emulates expert diagnostics by iteratively testing hypotheses in real-time until a root cause is confirmed (\cref{sec:online_diagnosis}). Figure~\ref{fig:Incident Mitigation Pipeline} illustrates the multi-agent architecture of \SYS in the online phase.

\subsection{Offline: Knowledge Consolidation}
\label{sec:offline_learning}
To address the critical gap of missing domain knowledge in off-the-shelf LLMs (Ch3), \SYS constructs three types of knowledge from on-call experience, including  
(1) Historical incident database: enables semantic failure pattern matching (\cref{sec:pipeline1}) in pipeline \#1;
(2) Hierarchical taxonomy: organizes the root cause labels for structured and staged reasoning (\cref{sec:pipeline2}) in pipeline \#2;
(3) Domain-specific rulebase: encapsulates expert knowledge as reusable prompt templates to guide both offline and all online decisions throughout pipelines \#1, 2, and 3.
\begin{algorithm}[t]
    \footnotesize
    \caption{Hierarchical Taxonomy Construction}\label{algo:taxonomy_construction}
    \KwIn{Incident dataset $D$;}
    \KwOut{Empty Taxonomy $T$;}
    
    \SetKwFunction{Diagnosis}{Diagnosis}  %
    \SetKwProg{Fn}{Function}{:}{\KwRet}  %

    \SetKwFunction{HierarchicalTaxonomyConstruction}{TaxonomyConstruction}  %
    \Fn{\HierarchicalTaxonomyConstruction{$D$}}{
        \ForEach{$incident$, $root\_cause \in D$}{
            \textcolor{blue}{\# [{\it LLM}] Classify root cause by taxonomy and semantics}\;
            $status \gets \texttt{LLM.ClassifyCause}(T, root\_cause)$ 
    
            \If{$status$ == \texttt{"EXIST"}} {
                \Comment{Existing taxonomy node: Link incident ID}

                $node = \texttt{LLM.FindSemanticNode}(root\_cause)$ 
                
                $node.incident\_ids.\texttt{append}(incident.{id})$
 }
            \ElseIf{$status$ == \texttt{"NEW"}} {
                \Comment{Novel cause: Extend taxonomy with new node}

                $node = T.\texttt{CreateNode}(incident, root\_cause)$

                $T.insert(node)$
 }
            \ElseIf{$status$ == \texttt{"NONE"}} {
                \Comment{Ambiguous cause: Skip processing}
                \texttt{Continue} 
 }
 }
        \ForEach{$node \in T$}{
            \Comment{Gather all incidents sharing the same root cause}
            $info = \texttt{GatherIncidents}(node.incident\_ids)$
            
            \textcolor{blue}{\# [{\it LLM}] Generate detailed description for the root cause}\;
            $node.description = {\texttt{LLM.Synthesis}}(info)$
    
            \textcolor{blue}{\# [{\it Manual}] Assign verification tools to the root cause}\;
            $node.verification = \texttt{AssignVerification(tools)}$
 }
        \Return $T$ \Comment{Return the constructed taxonomy}

 }

\end{algorithm}

{\bf Incident Gathering and Labeling.} 
Our methodology for creating a structured knowledge base begins with gathering and labeling historical incidents from Microsoft Azure. Each incident record consists of two primary components:
{\bf 1. Incident Description}: Contains technical artifacts reported by users, such as symptom observations, execution stack traces, and preliminary cause analyses.
{\bf 2. Postmortem Discussion}: Documents the iterative human investigation process, including multi-team discussions, hypothesis verification, and root cause confirmation.
We synthesize these components as input for an LLM, which is tasked with generating a structured, hierarchical root cause label.
This output label is formalized as a triplet:  {\it <main\_category>. <sub\_category>.<detailed\_error\_msg>}, and use prompt engineering to restrict the {\it main\_category} into six fixed categories:
\texttt{GPU, System Software, Interconnect \& Networking, Framework \& Library, User, and Other}. These categories are chosen based on our debugging experience and align with the classifications of other works \cite{deng2025minder}, while the {\it sub\_category} and {\it detailed\_error\_msg} fields are dynamically generated by the LLM to capture granular failure context.
Detailed definitions of these categories are provided in \cref{sec:taxonomy}.

{\bf Historical Incident Database Construction.}
The historical incident database is constructed to enable the rapid diagnosis of recurring issues through semantic pattern matching. The construction process involves two primary steps: {\bf semantic embedding} and {\bf efficient indexing}.
First, we use pre-trained embedding models~\cite{bge_embedding} to vectorize each incident description, encoding contextual semantics into a continuous vector space. Thus, geometric relationships (e.g., cosine distance) can quantify the semantic similarity of an incident.
Second, these vectorized descriptions are stored as index items, with their corresponding root cause labels attached as metadata.
During diagnosis, \SYS computes semantic similarity scores between the target incident and historical incident vectors, retrieves the top-K nearest neighbors using an approximate nearest neighbor (ANN) retrieval, and formulates hypotheses by aggregating the root cause labels (metadata) from these retrieved neighbors.

{\bf Hierarchical Taxonomy Construction.}
The incident taxonomy organizes root cause labels hierarchically, which is a common practice in cloud services~\cite{dogga2023autoarts}. This structure improves maintainability by localizing updates to specific branches without affecting the whole framework. However, manual construction typically demands multi-person-year effort~\cite{dogga2023autoarts} (expert elicitation, hierarchy validation, and alignment with evolving services), which does not scale.

To address scalability issues, we propose a semi-automated framework that leverages LLMs to derive the taxonomy directly from unstructured incident records. As outlined in Algorithm~\ref{algo:taxonomy_construction}, the workflow has two sequential phases: {\bf initial structure generation} and {\bf expert knowledge enrichment}. Starting with an empty taxonomy $T$ and the incident set $D$, Phase 1 (lines 2–12) iterates over all the incidents to construct the initial taxonomy.
For each root-cause label, the LLM determines whether and where its root-cause label should appear in $T$ (line 4). Given the triplet-format label and the current taxonomy context, the model classifies the label as new, existing, or none.
Based on this classification, it either creates a new node in $T$, links the incident to an existing node, or skips the label. 
This iterative process ensures the taxonomy remains comprehensive, flexible, and adaptable to emerging failure patterns.
%

Phase 2 (lines 13–19) enriches the taxonomy structure with actionable diagnostics. We first group incidents by root cause and use LLMs to synthesize concise descriptions for each node. Second, experts then assign verification steps per node using industry-standard tools, including SuperBench~\cite{xiong2024superbench}, NVIDIA DCGM~\cite{DCGM}, NCCL-test~\cite{nccl_tests}, dmesg~\cite{dmesg}, Azure Health Check~\cite{azure-health-checks}, etc. Notably, user-related incidents are excluded from this step, as they are typically difficult to diagnose with automated verification tools. This semi-automated workflow substantially reduces manual effort, allowing experts to focus on high-value tasks like validating new failure types and assigning context-specific verification logic.

A visualization of the taxonomy, built from one year of production incidents in Microsoft Azure, is provided in Figure~\ref{app:taxonomy}. 
 The taxonomy has three levels: 6 main categories, 28 subcategories, and 97 detailed error categories. Main and detailed categories include actionable verification steps for diagnosis, whereas intermediate subcategories serve only as organizational groupings.



{\bf Domain-specific Rulebase.}
To enhance incident diagnosis, we empirically encapsulate a set of text-based rules derived from on-call experience.
These rules are formulated as natural language statements that incorporate system configurations, internal software descriptions, common error categories, and their corresponding explanations.
The rulebase serves two critical functions:
(1) During offline processing, it guides automated labeling and taxonomy generation;
(2) For online diagnosis, it provides guidance for the Planning Agent's hypothesis generation during real-time troubleshooting.
In \SYS implementation, the rulebase is embedded into the prompt templates for both offline and online phases, ensuring consistent rule enforcement across all stages of the workflow.
\begin{figure}[t]
    \centering
    \includegraphics[width=0.85\linewidth]{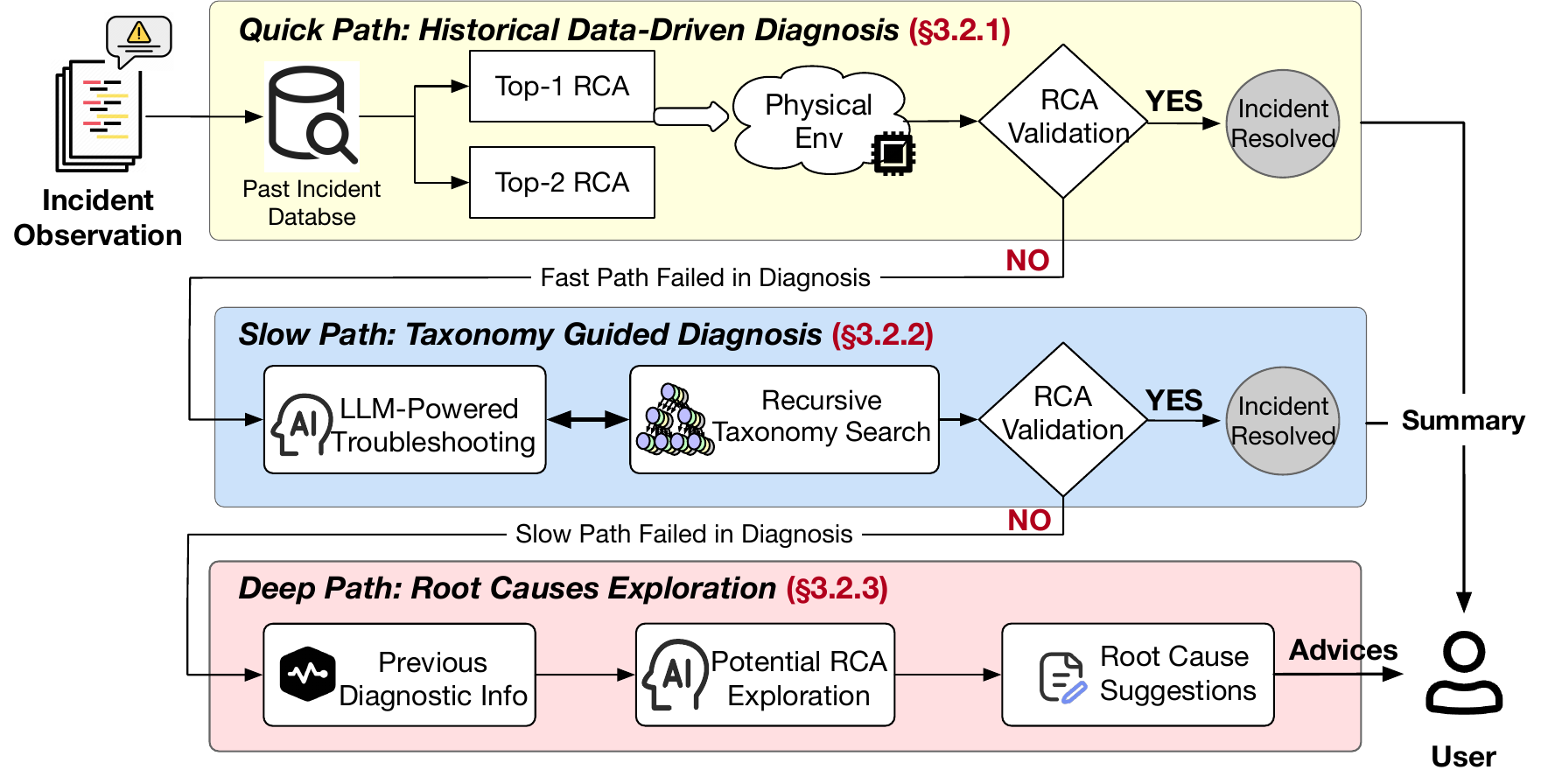}
    \vspace{-0.4cm}
    \caption{Tiered pipeline design for online incident diagnosis.}
    \label{fig:planning agent}
    \vspace{-0.4cm}
\end{figure}

\subsection{Online: Tiered Diagnosis Pipelines}
\label{sec:online_diagnosis}
    %
This entire online phase (Figure~\ref{fig:planning agent}) is architected to overcome key diagnostic challenges. To address the lack of iterative verification (Ch1), it employs a multi-agent framework that mimics expert reasoning through a dynamic cycle of hypothesis generation, evidence collection, and validation. To tackle unreliable semantic retrieval (Ch2), it uses a hierarchical pipeline that escalates from quick, pattern-based diagnosis Pipelines~\#1 (\cref{sec:pipeline1})  to systematic investigation in Pipelines~\#2 and \#3 (\cref{sec:pipeline2} and \cref{sec:pipeline3}) when simple matching is insufficient.

\subsubsection{Quick Path: Historical Data-Driven Diagnosis}
\label{sec:pipeline1}
The first diagnostic pipeline resolves recurring incidents through failure pattern matching, as illustrated in Figure~\ref{fig:planning agent}. The input is the incident description. The workflow begins with vectorizing the description using the BGE embedding model~\cite{bge_embedding} from the offline phase. This vector queries the pre-built historical incident database to retrieve the top five most similar historical incidents. 
These candidates are then processed by an LLM reranker~\cite{qin2023large}, which selects up to two most relevant cases based on textual similarity and incident context. The root causes of the selected cases become diagnostic hypotheses. 
 Each hypothesis is represented as a triplet-format root cause label. 
Subsequently, the execution agent executes the corresponding scripts and benchmarks for the specific root cause hypothesis and collects the results.
%
Then the reflection agent validates the hypotheses based on the collecting evidences.
When hypotheses are confirmed, such as the results indicating an obvious GPU fault or network degradation, the conclusion agent compiles a comprehensive diagnostic report for users. If no hypothesis is validated, the planning agent activates the taxonomy-guided diagnosis pipeline stage for further investigation.


\begin{algorithm}[t]
    \footnotesize
    \caption{Taxonomy-Guided Diagnosis}\label{algo:taxonomy_diagnosis}
    \KwIn{Incident description $I$, Pre-built taxonomy $T$;}
    \KwOut{Validated root cause $R$ or \texttt{None};}
    
    \SetKwFunction{Diagnosis}{Diagnosis}  %
    \SetKwProg{Fn}{Function}{:}{\KwRet}  %
    \Fn{\Diagnosis{$traversal\_node, M$}}{
        \textcolor{blue}{\# [{\it Reflection Agent}] Goal Found \& Backtracking}\;
        \If{{\tt IsLeafNode}($traversal\_node$)}{
            \If{{\tt LLM.Validation}($traversal\_node$)}{ 
                \KwRet $[traversal\_node]$ \Comment{Valid root cause found}
 }
            \KwRet \texttt{[]} \Comment{Backtrack to alternative paths}
 }
        
        \textcolor{blue}{\# [{\it Planning Agent}] State Expansion\& Branch Pruning}\;
        $sub\_categories \gets \texttt{LLM.Rank}(traversal\_node, M)$\;
    
        \textcolor{blue}{\# [{\it Execution Agent}] Environment Interactions} \;
        \ForEach{$sub\_category \in sub\_categories$}{
            $V \gets \texttt{EnvInteraction}(sub\_category)$; \Comment{Collect verification results}
            $M \gets \texttt{UpdateMemory}(M, sub\_category, V)$\;
            $R.\texttt{extend}(\texttt{Diagnosis}(sub\_category, M))$; \Comment{Recursive Exploration}
            $M \gets \texttt{ClearMemory}(M, sub\_category, V)$\;
 }
    
        \KwRet $R$
 }
    \SetKwFunction{TaxonomyGuidedDiagnosis}{TaxonomyGuidedDiagnosis}  %
    \Fn{\TaxonomyGuidedDiagnosis{$I, T$}}{
        \textcolor{blue}{\# [\textit{State Initialization}]}\;
        $root\_pointer \gets root(T)$ \Comment{Initialize at taxonomy root}
        $M \gets \texttt{InitializeMemory}(I, T)$ \Comment{Store incident context in agent memory}
        \KwRet $\texttt{Diagnosis}(root\_pointer , M)$ \Comment{Call Diagnosis to get result}
 }
    
\end{algorithm}

\subsubsection{Slow Path: Taxonomy-Guided Diagnosis}
\label{sec:pipeline2}
\begin{figure}[t]
    \centering
    \includegraphics[width=0.73\linewidth]{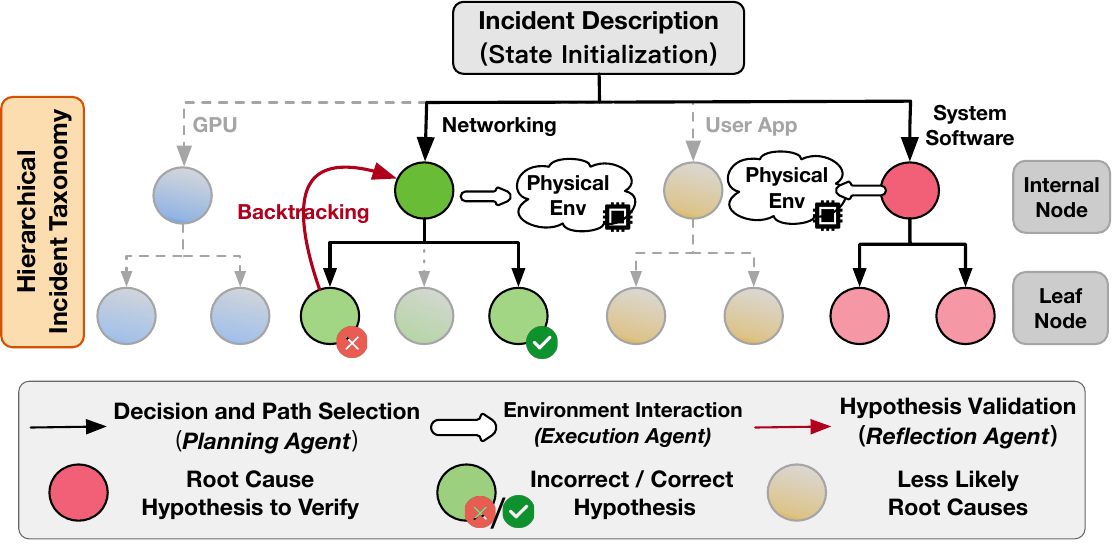}
    \vspace{-3mm}
    \caption{Illustration of recursive search in taxonomy-guided diagnosis.}
    \label{fig:recursive search}
    \vspace{-4mm}
\end{figure}

The slow path complements the quick path (\cref{sec:pipeline1}) by performing a comprehensive, taxonomy-guided diagnosis for more complex incidents. We formalize this process as a recursive taxonomy search algorithm (Algorithm~\ref{algo:taxonomy_diagnosis}) and exemplify its key procedures in Figure~\ref{fig:recursive search}. 

{\bf State Initialization.}
The process begins with two inputs: the incident description $I$ and the pre-built taxonomy $T$. As shown in Algorithm~\ref{algo:taxonomy_diagnosis} (lines 19-21), the planning agent initializes three key components: (1) a traversal pointer starting at the taxonomy's root, (2) a structured memory to store the diagnostic state (e.g., conversation history and current hypotheses), and (3) the incident description, which serves as a contextual anchor for further diagnosis.

{\bf Recursive Search.}
From the initialized root, memory, and incident description, the planning agent initiates a recursive exploration of the taxonomy, progressively refining the search space to identify the most probable root causes. The recursive search process integrates three tightly coupled phases: path selection (planning agent), environment interaction (execution agent), and backtracking (reflection agent), as illustrated in Figure~\ref{fig:recursive search}.

\textit{Path Selection (Planning Agent).}
This process is driven by the LLM's capabilities and guided by the taxonomy structure. At each search step, the planning agent selects up to three subcategories from the current traversal node, with selection criteria based on their relevance to both the incident description and verification results from the execution agent (line 8 in Algorithm~\ref{algo:taxonomy_diagnosis}). In \SYS, the searching strategy follows the Depth-first Search (DFS), prioritizing the most likely subcategories as initial hypotheses and invoking the execution agent for deeper exploration. When none of the subcategories are deemed relevant, the planning agent terminates the current branch and backtracks to evaluate alternative paths.

\textit{Environment Interaction (Execution Agent).}
For each subnode hypothesis generated by the planning agent, the execution agent interacts with the environment by running predefined verification scripts (lines 11-14 in Algorithm~\ref{algo:taxonomy_diagnosis}), collecting diagnostic data such as verification results and error logs. Nodes without associated scripts bypass execution, directly returning control to the planning agent. All collected evidences are stored at \SYS's memory, serving dual purposes: (1) guiding the planning agent's hypothesis refinement and (2) enabling the reflection agent's hypothesis validation. This dynamic interplay between taxonomy-driven logic and real-world interaction ensures accurate and systematic diagnoses.

\textit{Hypothesis Validation \& Backtracking (Reflection Agent).}
Upon reaching a leaf node, the reflection agent synthesizes accumulated verification evidence to validate the current hypothesis (lines 3-6 in Algorithm~\ref{algo:taxonomy_diagnosis}). It employs LLMs to determine whether the hypothesis is valid or should be rejected. If the hypothesis is confirmed, \SYS returns the validated root cause. Otherwise, the planning agent backtracks to the parent node and explores untested branches, ensuring no viable path is prematurely excluded.

Figure~\ref{fig:recursive search} exemplifies the slow path traversal through a hierarchical incident taxonomy. The process starts from an initial incident description. At each internal node, the {\it Planning Agent} acts as a navigator, selecting the most likely sub-paths for exploration. Following the chosen path, the {\it Execution Agent} gathers real-world feedback for each hypothesis. Upon reaching a leaf node, the {\it Reflection Agent} decides either confirming the hypothesis as the root cause or triggering a backtracking step to explore alternative branches.



\subsubsection{Deep Path: Root Cause Exploration}
\label{sec:pipeline3}
The first two pipelines are designed to diagnose incidents within predefined categories via pattern matching and structured reasoning based on histories data. However, evolving  AI infrastructure services and the increasing complexity of AI workloads make it challenging to cover all possible root causes in the taxonomy. To address this limitation, we designed a root cause exploration pipeline as a last resort for diagnosing incidents not covered by the first two pipeline stages.

This pipeline does not run extra scripts or tools, but rather collects verification results from the execution agent from the first two pipelines, using them as context to generate additional potential root causes. 
If all suggestions prove ineffective, the incident is escalated to the infrastructure support team for specialized investigation. 
In this case, a conclusion agent automatically compiles a comprehensive diagnostic report, detailing all actions taken by \SYS, to expedite the support team's understanding and reduce further investigation time.
\section{Implementation}
\label{sec:implementation}

We have developed \SYS using Python, totaling $\sim$5,200 lines of code (LoC). 
For historical incident retrieval, we employ Llama-index~\cite{llama_index} for data storage and utilize the BGE embedding model~\cite{bge_embedding} and a customized LLM-based reranker~\cite{qin2023large} to enhance retrieval accuracy.
For taxonomy management, we use anytree library~\cite{anytree} to store the taxonomy in JSON format. 
In our taxonomy generation framework, we use prompt engineering to restrict the <main\_category> into six fixed categories: \texttt{GPU,System Software,Interconnect \& Networking, Framework \& Library,User,and Other}, while the <sub\_category> and <detailed\_error\_msg> are generated by LLM.
%
\SYS also allows users to extend the taxonomy by adding new categories. Users are required to provide labels, detailed error descriptions, and corresponding verification scripts to facilitate accurate diagnosis. 

\section{Evaluation}
\label{sec:evaluation}

We aim to systematically evaluate the \SYS design by answering the following four research questions (RQs) and presenting real-world case studies (\cref{sec:deep_dive}) to illustrate how it can disentangle interdependent symptoms.

\noindent $\bullet$ RQ1: How does \SYS compare to baseline methods in terms of diagnostic performance across different incident categories (\cref{sec:evaluation overall})?

\noindent $\bullet$ RQ2: How do individual components contribute to \SYS's overall performance (\cref{sec:evaluation overall})?

\noindent $\bullet$ RQ3: What is the computational efficiency and runtime overhead of \SYS during online diagnosis (\cref{sec:evaluation_cost})?

\noindent $\bullet$ RQ4: What is the offline overhead for constructing \SYS's knowledge base (\cref{sec:evaluation_cost})?


\subsection{Experiment Setup}
\noindent {\bf Incident Ticket Datasets.}
\label{sec:data_collection}
We collect real-world incident tickets from production-scale GPU clusters in Microsoft Azure, spanning from 2023-04 to 2024-03. 
For focusing on AI workload incidents, we filtered the incident tickets using keywords such as \texttt{GPU}, \texttt{CUDA}, \texttt{NCCL}, \texttt{PyTorch}, \texttt{TensorFlow}, etc. 
For scientific validity, tickets with no clear root causes were manually excluded. This process resulted in a dataset of 778 incidents. Each incident ticket includes a detailed description, identified root causes, on-call engineers' discussion histories, and resolution steps.
Following the chronological order, the latter 25\% of the incidents (208) were set aside as the test set, while the former 75\% were used to construct the historical incident database and taxonomy.
The test set covers a diverse range of incident categories: 115 GPU-related incidents, 53 system software issues, 22 interconnect and networking problems, 16 user application incidents, and 2 framework and library-related incidents.

{\bf Baselines.} We compare \SYS against the following four baselines:

\noindent $\bullet$ {\bf \textit{RCACopilot}}: A SOTA LLM-based root cause analysis design for cloud services~\cite{RCACopilot}. Our implementation does not include the information collection stage mentioned in RCACopilot, as it is specifically tailored for email services.

\noindent $\bullet$ {\bf \textit{Taxonomy-Guided Diagnosis (TGD)}}: This baseline only uses the slow path for diagnosis, taking the incident description as input and using the LLM to traverse the taxonomy for diagnosis (\cref{sec:pipeline2}).

\noindent $\bullet$ {\bf \textit{Direct Incident Diagnosis with o1 (DID-o1)}}: This baseline leverages the advanced o1-preview model~\cite{azure_oai_o1} for direct incident diagnosis. By employing the ``reasoning'' model, it accepts incident descriptions as input and offers users the most likely root cause label.

\noindent $\bullet$ {\bf \textit{Comprehensive Verification Diagnosis (CVD)}}: This baseline adopts a non-selective, brute-force approach. For any given incident, it sequentially executes all available verification scripts to gather comprehensive diagnostic data. The LLM then receives the incident description along with the full output of these scripts, and makes a final judgment on the root cause based on this complete context.

We use GPT-4o~\cite{OpenAIGPT-4o} from the Azure OpenAI service~\cite{azure_oai_service} as the default LLM for all baselines, except for DID-o1.

{\bf Metrics.}
We evaluate \SYS using Precision, Recall, and Micro/Macro F1-scores, comparing its performance against baseline methods.
The Micro F1-score aggregates performance across all classes, weighting each sample equally, while the Macro F1-score evaluates performance on a per-class basis, treating all classes equally regardless of their size. 
Additionally, we assess its efficiency in terms of verification time and LLM invocation overhead.

{\bf Fidelity of Incident Categories.}
To ensure label accuracy in the offline stage, we perform a rigorous fidelity check on the taxonomy labels used throughout TSGuard's knowledge base.
Specifically, three senior on-call engineers (OCEs), each with over three years of production experience, independently validated the taxonomy labels of 50 randomly sampled test set tickets against post-mortem ground truth---i.e., the verified root cause documented in the resolution report.
Each OCE assessed whether the LLM-assigned taxonomy label correctly matched the ground truth root cause for each incident.
Out of the 50 sampled incidents, 48 labels were unanimously confirmed as correct by all three OCEs, yielding a 96\% accuracy rate (48/50).
The two mismatched cases involved ambiguous symptom descriptions where the LLM selected a closely related but incorrect sub-category.
This validation confirms that the taxonomy is both \emph{complete} (covering production failure modes) and \emph{unambiguous} (labels are semantically distinct enough for independent annotators to agree), supporting its reliability for both knowledge construction and evaluation.

\begin{figure*}[t]
        \centering
    \includegraphics[width=0.7\linewidth]{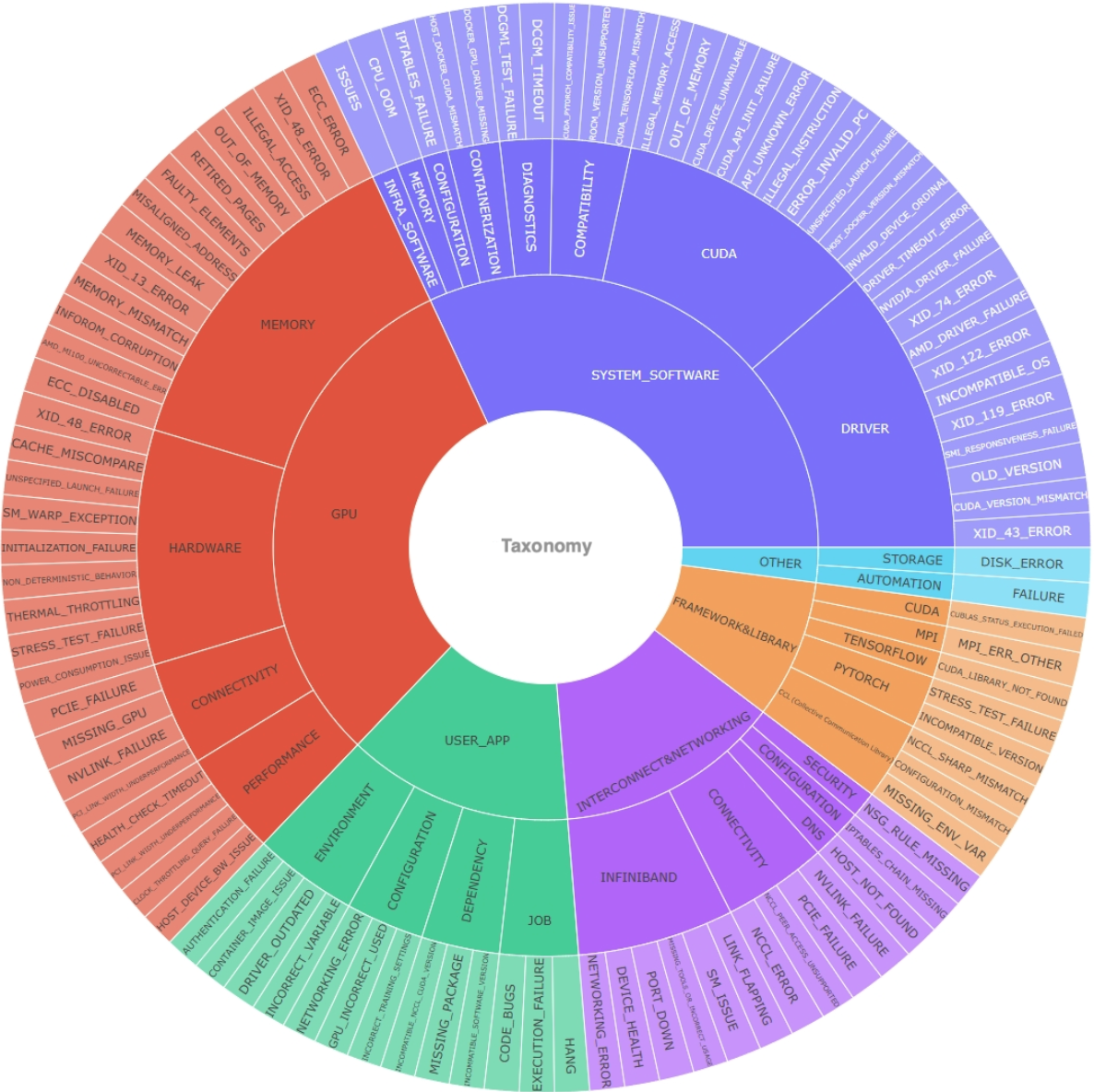}
        \caption{Visualization of the incident taxonomy.}
        \label{app:taxonomy}
\end{figure*}

{\bf Taxonomy Visualization.}
\label{sec:taxonomy}
The hierarchical incident taxonomy is visualized in Figure~\ref{app:taxonomy}.
Note that this taxonomy is generated by LLMs based on the historical incident records and expert knowledge. It still needs to be validated by domain experts.
The main categories in the taxonomy include:
\textbf{GPU} (hardware-related),
\textbf{System Software} (driver, CUDA, infrastructure),
\textbf{Interconnect \& Networking} (network and high-speed interconnects),
\textbf{Framework \& Library} (AI frameworks),
\textbf{User Application} (user errors, configuration conflicts), and
\textbf{Other}.

\begin{figure}[t]
    \centering
    \includegraphics[width=0.7\linewidth]{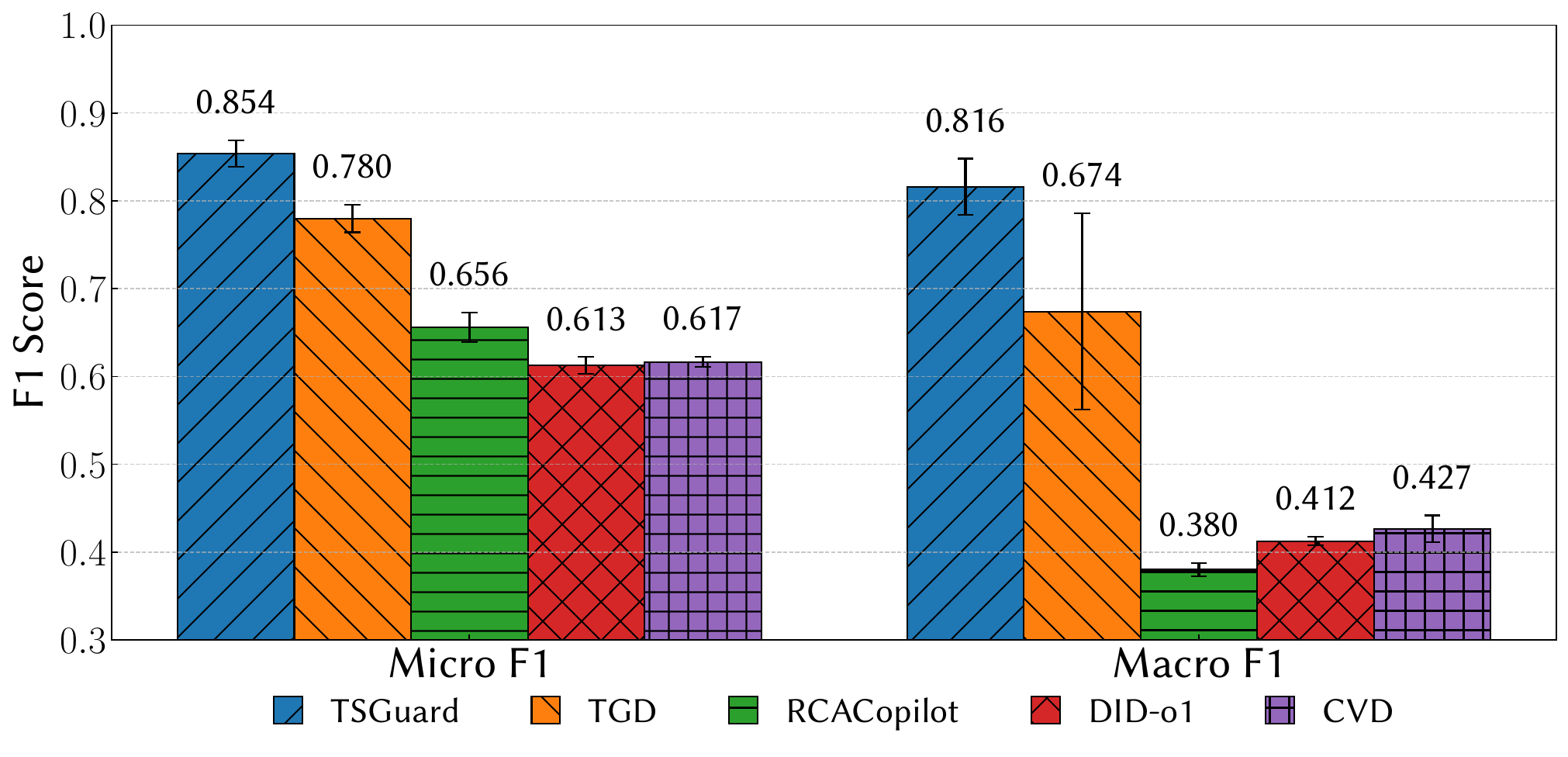}
    \vspace{-1em}
    \caption{Comparison of Micro F1 and Macro F1 scores between \SYS and baseline methods. Error bars represent the standard deviation across experimental runs.}
    \label{fig:micro_macro_f1}
\end{figure}

\begin{figure}[t]
    \centering
    \includegraphics[width=0.8\linewidth]{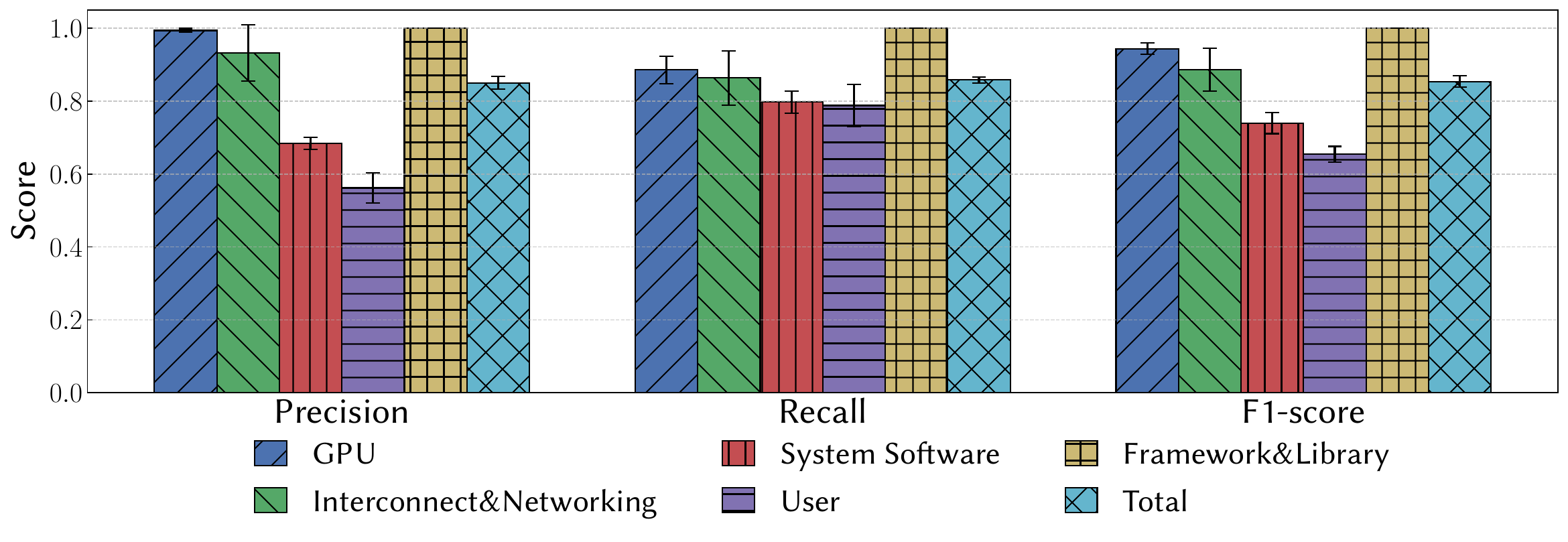}
    \caption{The Precision, Recall, and Micro F1 Score of \SYS across incident categories}
    \label{fig:breakdown_categories_performance}
\end{figure}

\subsection{Overall Performance}
\label{sec:evaluation overall}
\textbf{Effectiveness of Diagnosis.}
We evaluate the overall performance of \SYS and all baseline methods using the Micro and Macro F1-scores. 
The error bars represent the standard deviation of F1 scores across multiple experimental runs.

As shown in Figure~\ref{fig:micro_macro_f1}, \SYS consistently outperforms all baseline methods in both Micro F1 and Macro F1 scores, demonstrating its superior diagnostic capability. Specifically, \SYS achieves the highest Micro F1 score of 0.854, significantly surpassing the second-best method, TGD, which achieves a score of 0.780. 
In terms of the Macro F1 score, the gap in \SYS (0.816) and remaining baseline methods is even more pronounced, with RCACopilot, DID-o1, and CVD scoring 0.380, 0.424, and 0.427, respectively.
Notably, despite the integration of comprehensive verification in CVD, its diagnostic accuracy remains limited (0.427 Macro-F1). By checking the results of CVD, we find that the LLM is easy to be misled by trivial testing jitter results and ignores the pivotal benchmarks, indicating that structured reasoning and feedback mechanisms are essential for prioritizing root causes within hierarchical taxonomies.
Moreover, we conduct each experiment five times and find that \SYS exhibits smaller error bars compared to most of the baseline methods in Micro F1. While for Macro F1, the error bars of \SYS are slightly larger than RCACopilot, DID-o1, and CVD, this is due to certain categories having fewer samples, leading to higher variance in the evaluation results.
Overall, these results underscore the effectiveness of \SYS in accurately predicting root cause categories while maintaining stability under varying conditions.

\begin{table}[t]
    \centering
    \caption{Performance contribution factors analysis in \SYS.}
    \vspace{-1em}
    \label{table:breakdown_performance}
    \normalsize
    \renewcommand{\arraystretch}{0.85} %
    \resizebox{0.7\columnwidth}{!}{
        \begin{tabular}{c|cc}
            \hline
            \SYS                                                                                      & \begin{tabular}[c]{@{}c@{}}Average Resolved\\ Incidents\end{tabular} & \begin{tabular}[c]{@{}c@{}}Cumulative\\ Percentage\end{tabular} \\ \hline
            \begin{tabular}[c]{@{}c@{}}Historical Data Driven\\ \textbf{(Pipeline \#1)}\end{tabular}                          & 107.0                                                               & 51.4\%                                                           \\
            \begin{tabular}[c]{@{}c@{}}Historical + Taxonomy-Guided\\ \textbf{(Pipeline \#1 + \#2)}\end{tabular}              & 174.9                                                               & 84.0\%                                                           \\
            \begin{tabular}[c]{@{}c@{}}Historical + Taxonomy + Exploration\\ \textbf{(Pipeline \#1 + \#2 + \#3)}\end{tabular} & 179.6                                                               & 86.3\%                                                           \\ \hline
            \end{tabular}
 }
\end{table}

\begin{table}[t]
    \centering
    \caption{Performance of \SYS on unseen incidents.}
    \vspace{-1em}

        \label{table:unseen_incidents_perfromance}
        \normalsize
        \renewcommand{\arraystretch}{1} %
        \resizebox{0.7\columnwidth}{!}{
            \begin{tabular}{c|ccc}
                \hline
 Tickets                 & \begin{tabular}[c]{@{}c@{}}Removed Taxonomy\\ Label\end{tabular}                             & \begin{tabular}[c]{@{}c@{}}Accuracy After \\ Removal\end{tabular} & \begin{tabular}[c]{@{}c@{}}Accuracy Before \\ Removal\end{tabular} \\ \hline
                \multirow{2}{*}{Simple} & \begin{tabular}[c]{@{}c@{}}GPU.MEMORY.\\ InfoROM\_Corruption\end{tabular}                    & 75\%                                                              & 100\%                                                              \\ \cline{2-4}
                                        & \begin{tabular}[c]{@{}c@{}}SYSTEM\_SOFTWARE.CUDA.\\ Illegal\_Mem\_Access\end{tabular}        & 93.8\%                                                           & 100\%                                                              \\ \hline
 Complex                 & \begin{tabular}[c]{@{}c@{}}SYSTEM\_SOFTWARE.CUDA.\\ Host\_VM\_Version\_Mismatch\end{tabular} & 0\%                                                               & 42.5\%                                                             \\ \hline
                \end{tabular}
 }

\end{table}

{\bf Performance Across Incident Categories.}
To comprehensively assess \SYS's effectiveness, we conduct a category-wise performance analysis as shown in Figure~\ref{fig:breakdown_categories_performance}.
While \SYS achieves strong Precision, Recall, and Micro-F1 scores in most categories, its performance significantly degrades for user-related incidents. 
This limitation stems from inherent biases in the incident collection methodology: these collected ``user-related'' incidents predominantly represent errors triggered by internal infrastructure components (e.g., deployment scripts or configuration APIs) during user operations. Although these incidents contain infrastructure-originated error messages, their root causes are actually user-side operational errors. This semantic mismatch between observed symptoms (infrastructure errors) and underlying causes (user actions) causes \SYS to incorrectly prioritize infrastructure-level root causes, thereby reducing its effectiveness on user-related cases.


{\bf In-depth Component Performance Analysis.}
%
To quantify component contributions, we perform a granular evaluation of the tiered pipeline architecture. As detailed in Table~\ref{table:breakdown_performance}, the standalone historical data-driven diagnosis component (Pipeline \#1) addresses 51.4\% of incidents, demonstrating the foundational value of leveraging past incident data for new incidents. By augmenting this with taxonomy-guided diagnosis (Pipeline \#1 + \#2), we observe a 32.6 percentage point increase in resolution rate (from 51.4\% to 84.0\%), validating the indispensable role of structured reasoning in improving diagnostic precision.
The full \SYS design (Pipeline \#1 + \#2 + \#3) further improves performance to 86.3\%, with the 2.3-point gain attributable to the exploration stage.
These results validate the cumulative efficacy of our tiered design, where each component provides complementary enhancements to \SYS's overall diagnostic capability.


{\bf Handling Unseen Incidents.}
To assess \SYS's robustness against out-of-distribution incidents, we simulate label scarcity scenarios by iteratively removing critical labels from the taxonomy and measuring diagnostic accuracy degradation. Results in Table~\ref{table:unseen_incidents_perfromance} reveal a bimodal behavior:
Simple cases (e.g., \texttt{infoROM corruption}, \texttt{illegal memory access}) maintain 75\%–93.75\% accuracy upon post-label removal, indicating resilience through the exploration diagnostic path.
Complex interdependent failures (e.g., \texttt{Host VM CUDA version mismatch}) suffer catastrophic accuracy drops (42.5\% $\to$ 0\%), exposing reliance on explicit taxonomy guidance for multi-fault reasoning.
This phenomenon reveals a fundamental trade-off in \SYS's design: although the root cause exploration phase enables robust performance on routine incidents, the absence of taxonomic anchors critically limits its ability to resolve domain-specific incidents. Nevertheless, most of the incidents were repetitive during our one-year observation, and no significant new incidents occurred in practice.

\begin{figure}[t]
    \centering
    \includegraphics[width=0.75\linewidth]{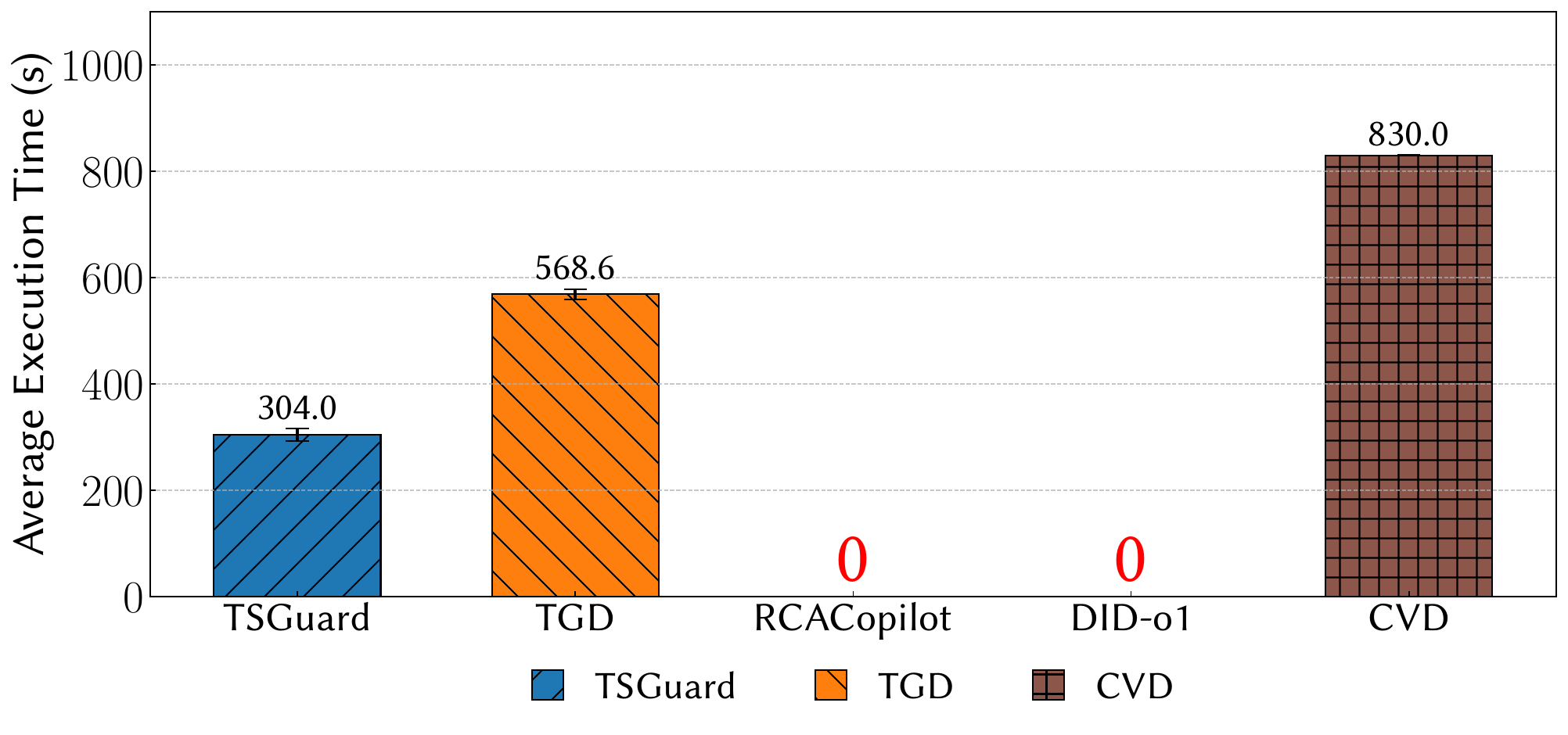}
    \vspace{-1.5em}
    \caption{Average verification time across different diagnostic methods on one single machine.}
    \vspace{-1.5em}

    \label{fig:verification_time}
\end{figure}


\begin{figure}[t]
    \centering
    \includegraphics[width=0.75\linewidth]{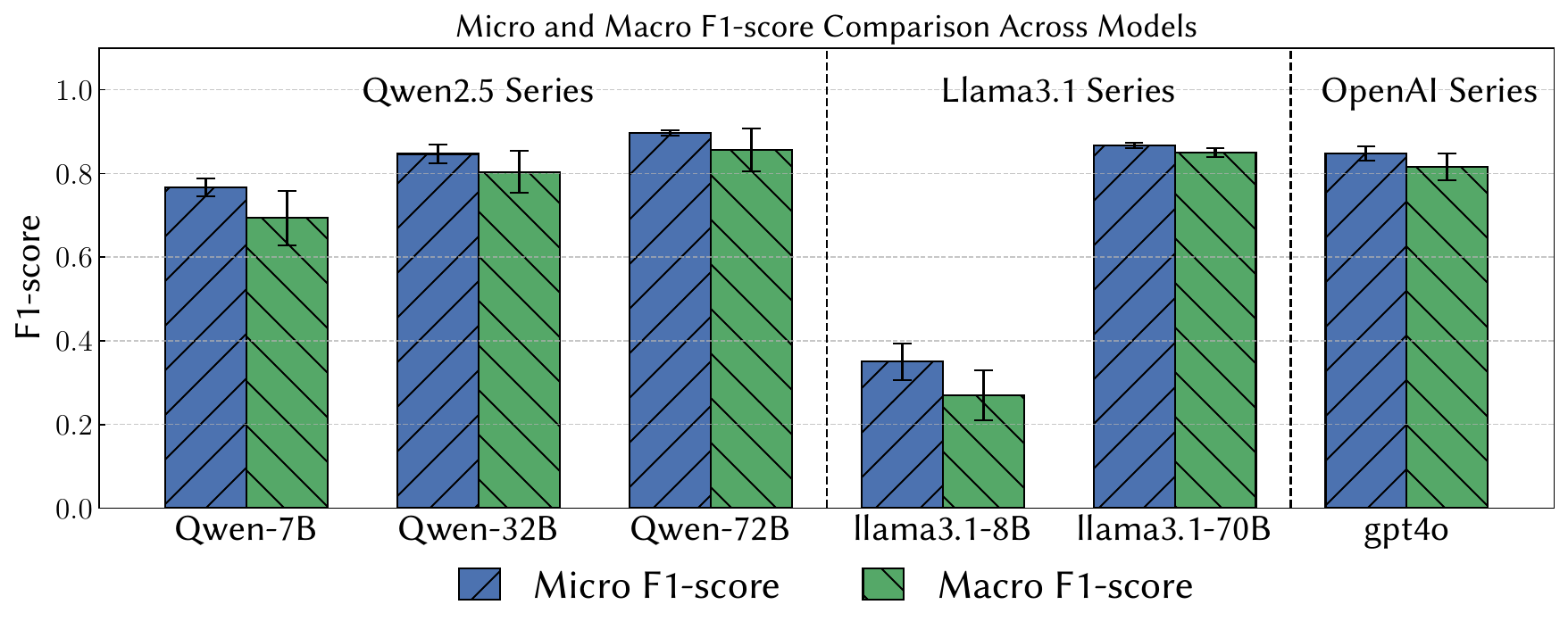}
    \vspace{-1em}
    \caption{The F1 scores of \SYS when using different LLMs.}
    \vspace{-1em}
    \label{fig:various_llm_performance}
\end{figure}

{\bf Impact of LLMs on Diagnostic Accuracy.}
 We also change the LLMs in \SYS to evaluate the impact of different LLMs on diagnostic accuracy. We use the most recent open-source LLMs: Llama3.1~\cite{dubey2024llama} (8B~\cite{llama3.1-8b} and 70B~\cite{llama3.1-70b}) and Qwen2.5~\cite{yang2024qwen2} (7B~\cite{qwen2.5-7b}, 32B~\cite{qwen2.5-32b}, and 72B~\cite{qwen2.5-72b}), measure the performance and compare them with the default GPT-4o\cite{OpenAIGPT-4o} in \SYS. Note that, we use the GPTQ~\cite{frantar2022gptq} quantization version and locally deploy them with SGLang~\cite{zheng2023efficiently}.

 Figure~\ref{fig:various_llm_performance} compares the Micro and Macro F1-scores across these models. The results show that the Qwen2.5 series consistently achieves higher scores compared to other models, with the Qwen-72B demonstrating the best overall performance. The Llama3.1 series, while competitive, shows a noticeable gap in the Macro F1-score, particularly for the smaller 8B model, indicating that model size and architecture significantly impact diagnostic accuracy. These results show that the \SYS design can also be locally deployed with open-source LLMs and larger models tend to provide better diagnostic accuracy.

\subsection{Cost Efficiency Analysis}
\label{sec:evaluation_cost}
{\bf Verification Time Benchmarking.}
To quantify operational efficiency, we measure end-to-end verification latency, defined as the total time from diagnostic trigger to actionable result delivery, including script execution and telemetry collection phases. As shown in Figure~\ref{fig:verification_time}, \SYS achieves a 63.4\% reduction in latency compared to the CVD baseline (304.0s vs. 830.0s), and outperforms TGD by 46.5\% (304.0s vs. 568.6s). RCACopilot and DID-o1 are excluded from this benchmark as they lack automated verification modules. This efficiency gain stems from \SYS's incremental verification strategy (Algorithm~\ref{algo:taxonomy_diagnosis}), which prioritizes high-probability fault candidates identified during tiered diagnosis, thereby avoiding exhaustive checks. 

{\bf Cost of Constructing Domain-specific Knowledge Base.}
\label{sec:offline_cost}
We quantify offline overhead across three processes: (1) LLM-based incident labeling, (2) embedding/indexing for the historical database, and (3) taxonomy construction. We focus on model invocation costs (other compute is negligible), using GPT-4o~\cite{OpenAIGPT-4o} for labeling/taxonomy and a locally deployed BGE-large~\cite{bge_embedding} for indexing. Table~\ref{table:offline_overhead} reports invocations, tokens, processing time, and cost.

The analysis breaks down as follows:
First, ticket labeling required 775 GPT-4o invocations, averaging 1530.2 input and 120.8 output tokens per call, which took 60.8 seconds and cost \$7.79.
Second, indexing the historical incident database involved 570 BGE invocations, a process that took 25.4 seconds and incurred negligible cost due to local model deployment.
Third, taxonomy construction utilized 570 GPT-4o invocations, with an average of 2898.1 input and 142.8 output tokens per iteration, taking 448.4 seconds and costing \$9.89.
While labeling and indexing are parallelizable, taxonomy construction is inherently sequential, as each step depends on the previous one. This sequential nature explains its longer execution time. The resulting taxonomy is visualized in Figure~\ref{app:taxonomy}.

\begin{table}[t]
    \centering
    \caption{Cost of knowledge consolidation during offline processing.}
    \label{table:offline_overhead}
    \vspace{-0.3cm}
    \resizebox{0.75\columnwidth}{!}{
        \begin{tabular}{ccccccc}
            \hline
            \begin{tabular}[c]{@{}c@{}}Processing\\ Stage\end{tabular} & Model Used                                            & Invocations & \begin{tabular}[c]{@{}c@{}}Input \\ Tokens\end{tabular} & \begin{tabular}[c]{@{}c@{}}Output \\ Tokens\end{tabular} & \begin{tabular}[c]{@{}c@{}}Times\\ (Sec)\end{tabular} & \begin{tabular}[c]{@{}c@{}}Cost\\ (USD)\end{tabular}       \\ \hline
 Labeling         & GPT-4o~\cite{OpenAIGPT-4o}      & 775         & 1530.2                                                  & 120.8                                                    & 60.8                      & \$7.79 \\
 Indexing         & BGE-large~\cite{bge_embedding} & 570         & \xmark                                                        & \xmark                                                         & 25.4                      & \$0 \\
 Taxonomy         & GPT-4o~\cite{OpenAIGPT-4o}      & 570         & 2898.1                                                  & 142.8                                                    & 448.4                     & \$9.89 \\ \hline
            \end{tabular}
 }
 \vspace{-0.6cm}
\end{table}

\begin{table}[t]
    \centering

    \caption{Average LLM invocations, total input tokens, total output tokens, and price of API calls for each diagnostic method per incident.}
    \label{table:llm_invocations}
    \vspace{-0.5em}

    \resizebox{0.7\columnwidth}{!}{
        \begin{tabular}{cccccc}
            \hline
            \begin{tabular}[c]{@{}c@{}}Diagnostic \\ Method\end{tabular} & \begin{tabular}[c]{@{}c@{}}LLM \\ Model\end{tabular} & \begin{tabular}[c]{@{}c@{}}\# of LLM \\ Invocations\end{tabular} & \begin{tabular}[c]{@{}c@{}}Input\\ Tokens\end{tabular} & \begin{tabular}[c]{@{}c@{}}Output \\ Tokens\end{tabular} & \begin{tabular}[c]{@{}c@{}}Price\\ (USD)\end{tabular} \\ \hline
            \textbf{\SYS}                                 & \textbf{GPT-4o}~\cite{OpenAIGPT-4o}                                      & \textbf{10.2}                                                    & \textbf{31801.2}                                        & \textbf{2224.9}                                          & \textbf{\$0.102}                                      \\
 TGD                                                          & GPT-4o~\cite{OpenAIGPT-4o}                                               & 18.17                                                            & 82070.0                                                 & 4735.5                                                   & \$0.253                                               \\
 RCACopilot                                                   & GPT-4o~\cite{OpenAIGPT-4o}                                               & 1.13                                                             & 817.1                                                   & 109.6                                                    & \$0.003                                               \\
 DID-o1                                                       & o1-preview~\cite{azure_oai_o1}                                           & 1                                                                & 2877.6                                                  & 136.9                                                    & \$0.051                                               \\
 CVD                                                          & GPT-4o~\cite{OpenAIGPT-4o}                                               & 1                                                                & 8037.2                                                  & 160.5                                                    & \$0.022                                               \\ \hline
            \end{tabular}
 }
\vspace{-1em}
\end{table}

{\bf Cost of Online Diagnosis.}
We quantify the overhead for online diagnosis by analyzing the average number of LLM invocations, total input tokens, and total output tokens per incident over all baselines, as shown in Table~\ref{table:llm_invocations}.
Using Azure OpenAI Service's pricing model~\cite{Azure-openai-pricing}, we further convert these metrics to monetary costs, which reveals stark contrasts:
TGD emerges as the most expensive baseline (\$0.253/incident), 2.5× higher than \SYS. Lightweight methods (RCACopilot, DID-o1, CVD) operate at \$0.003 - \$0.051/incident, but sacrifice diagnostic accuracy (per Figure~\ref{fig:micro_macro_f1}).
\SYS falls between these two extremes, with a cost of \$0.102/incident. 
However, considering the higher Micro/Macro F1 score of \SYS, this overhead is justifiable. Moreover, this overhead can be further reduced using optimization techniques from existing compound LLM systems~\cite{tan2024teola, lin2024parrot}.



\subsection{Case Study with Real Incidents}
\label{sec:deep_dive}

This section presents real-world case studies to demonstrate how \SYS's iterative diagnosis and structured reasoning can disentangle interdependent symptoms. We also analyze the cases unresolved by \SYS, revealing their inherent complexity, which poses challenges even for human experts.

{\bf Incident \#1: NCCL Error: Connection Refused.}
Figure~\ref{fig:example_nvlink_error} shows an NCCL connection refused error that caused failures in distributed training. 
Due to insufficient information in the incident description, \SYS's first pipeline (\cref{sec:pipeline1}) failed to retrieve relevant past incidents.
Consequently, \SYS activated its taxonomy-guided diagnosis pipeline (\cref{sec:pipeline2}), visualized in Figure~\ref{fig:deep_dive_example}(a).
Based on the description, the planning agent initially hypothesized \texttt{Interconnect \& Networking}, \texttt{System Software}, and \texttt{User Applications} as potential root cause categories. A subsequent DFS search, starting from the most likely category, ultimately identified \texttt{NCCL Error} and \texttt{NVLink\_Failure} as the root causes.


{\bf Incident \#2: Uncorrectable ECC Error.}
The second example (Figure~\ref{fig:example_ecc_error}) involves an uncorrectable ECC error encountered during gpt2 pretraining. 
The raw incident report exhibited multi-modal symptoms: CUDA kernel failures, GPU memory ECC error, and PyTorch allocation fragmentation.
%
Consequently, using this incident summary as input, Pipeline\#1 (\cref{sec:pipeline1}) failed to retrieve relevant historical cases, as the incident description diluted the critical ECC error signals.
\SYS then switched to Pipeline \#2 (\cref{sec:pipeline2}) for deeper analysis, as shown in Figure~\ref{fig:deep_dive_example}(b). 
The planning agent first hypothesized potential causes: \texttt{GPU}, \texttt{Framework \& Library}, and \texttt{System Software}, ordered by likelihood.
Following a DFS approach, \SYS explored these possibilities and confirmed the root causes as \texttt{ECC Error}, \texttt{Page Retirement}, and \texttt{Xid 48 Error}.
The ECC error was the primary trigger, while the other two were its downstream effects.


{\bf Unsolved Incidents.}
Despite \SYS's strong diagnostic capabilities, some incidents remain challenging. We identified 28 failure cases ($\sim$13.5\%) that \SYS failed to resolve in a single run.
Their TTM analysis confirms their inherent complexity: while the median TTM for all incidents is 52.50 hours (mean 83.00 hours), these unresolved cases by \SYS had a median TTM of 122.56 hours (mean 164.39 hours). 
This significantly elevated TTM indicates that the incidents \SYS failed to resolve are, by nature, more complex and time-consuming even for human experts.

\begin{figure}[t]
    \vspace{-0.5em}
    \begin{tcolorbox}[title=Example 1 Incident Description - NVLink Error (Root Cause), 
 fonttitle=\scriptsize, colframe=black, boxrule=0.5pt,
 left=0pt, right=2pt, top=0pt, bottom=0pt, enhanced jigsaw]

 {\scriptsize Customer failed to deploy distributed job on xx due to rank-0 can't talk to peer node.
 They tried this for six times (or more) and the bad instance are all point to the same physical node.

 {\bf Sample user log:}
 {\tt [4] <job\_id>:684:1526 [0] include/socket.h:406 NCCL WARN Connect to ip\_addr<port> failed : Connection refused}

\textbf{OCE Summary:} All links to GPU1 are inactive. See pasted \texttt{nvlink -s} results.
 }
    \end{tcolorbox}
    \vspace{-1em}
    \captionof{figure}{Incident description of Example 1. The root cause is NVLink failure due to inactive links.}
    \label{fig:example_nvlink_error}
\end{figure}

\begin{figure}[t]
    \vspace{-0.7em}
    \begin{tcolorbox}[title=Example 2 Incident Description - ECC Error (Root Cause), 
 fonttitle=\scriptsize, colframe=black, boxrule=0.5pt,
 left=0pt, right=0pt, top=0pt, bottom=0pt, enhanced jigsaw]
 {\scriptsize
{\tt node-92: Traceback (most recent call last):}

{\tt node-92:   File "pretrain\_gpt2.py", line 227, in <module>}

{\tt node-92:     pretrain(}



{\tt ...}

{\tt node-92: RuntimeError: CUDA error: CUBLAS\_STATUS\_EXECUTION\_FAILED}

{\tt node-92:  when calling `cublasGemmEx(...)`}

{\tt node-92: terminate called after throwing an instance of 'c10::Error'}

{\tt node-92:   what():  CUDA error: uncorrectable ECC error encountered}

{\tt node-92: CUDA kernel errors might be asynchronously reported at some other API call.}

{\tt node-92: For debugging consider passing CUDA\_LAUNCH\_BLOCKING=1.}

{\tt node-92: Exception raised from c10\_cuda\_check\_implementation at}

{\tt ../c10/cuda/CUDAException.cpp:31 (most recent call first):}

{\tt frame \#0: c10::Error::Error... (omitted)}

{\textbf{OCE Summary}: GPU ECC health check failed. This issue was fixed by reboot.}
    
 }
    \end{tcolorbox}
    \vspace{-1em}
    \captionof{figure}{Incident description of Example 2. The root cause is an ECC Error due to a health check failure.}
    \vspace{-1em}
    \label{fig:example_ecc_error}
\end{figure}

\vspace{-0.28em}
\section{Discussion and Threats to Validity}
\label{sec:discussion}



{\bf Internal validity.}
We audited 50 randomly sampled test tickets and observed 96\% agreement between LLM and human annotations (Section~\ref{sec:evaluation}); the remaining 4\% and taxonomy triplet errors may still introduce noise into knowledge construction and evaluation. Our dataset comes from a single provider and is keyword-filtered (GPU/CUDA/NCCL), which may over-represent infrastructure incidents. We mitigate via per-category reporting, a fixed split, and repeated runs, but cannot eliminate the threat entirely.

{\bf External validity.}
Our empirical data comes from Microsoft Azure. While the \SYS methodology is provider-agnostic, the concrete taxonomy, rules, and scripts are influenced by hardware SKU, software stack, and operational processes. Porting to other clouds or on-prem clusters likely requires re-grounding the taxonomy and verification scripts. Likewise, incidents beyond GPU-centric AI workloads (e.g., CPU-only analytics, storage-heavy services) may require new categories/tools before comparable accuracy holds. 

{\bf Mitigation Suggestions.}
For safety, TSGuard's scope is strictly diagnosis, not mitigation, as allowing LLM-based systems to perform unsupervised mitigation poses significant risks. During an incident, users still need to wait for cloud provider OCEs to remediate system errors.

{\bf Limitations and Future Work.}
\SYS's multi-tiered approach inherently supports detecting unseen incidents: its third pipeline targets unknown root causes by analyzing accumulated verification evidence without relying on historical patterns.
The current performance gap in the ``User Error'' category stems from attempting to pinpoint specific user-side root causes, which is impractical due to privacy and permission boundaries.
As future work, we plan to establish ``Infrastructure Healthy'' as an explicit diagnostic conclusion---if \SYS verifies that hardware, network, and drivers are all functional, this reverse proof confirms the problem lies on the user side, still providing actionable guidance.

\section{Related Work}
\label{sec:related}

\begin{figure}[t]
    \centering
    \includegraphics[width=0.75\linewidth]{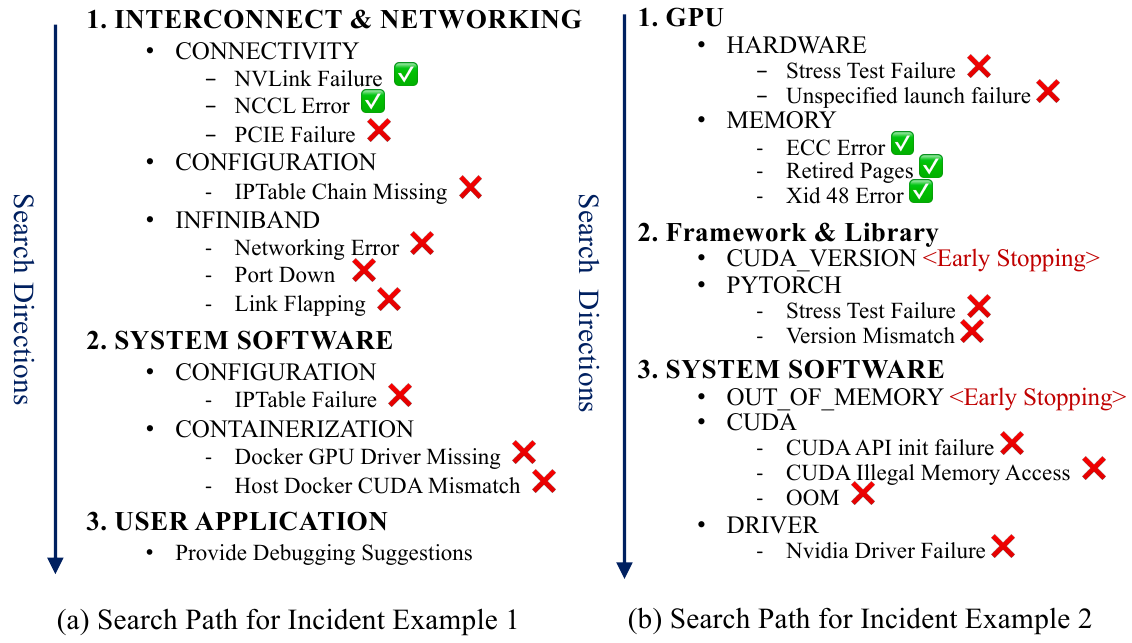}
    \vspace{-1em}
    \caption{Diagnosis visualization for the two examples in Figure~\ref{fig:example_nvlink_error} and~\ref{fig:example_ecc_error}. Green checkmarks (\cmark) and red crosses (\xmark) indicate the reflection agent's confirmation or rejection of hypotheses, while the red \texttt{<Early Stopping>} marker denotes branch pruning by the planning agent.}
    \label{fig:deep_dive_example}
    \vspace{-1.3em}
\end{figure}




Significant efforts have been dedicated to automated incident troubleshooting for cloud services, with a primary focus on reducing the burden on OCEs~\cite{RCACopilot, jin2023assess, dogga2023autoarts, ahmed2023recommending, zhang2024automated,liu2025opseval, zhang2024lm, jiang2024xpert,li2023exploring, an2024nissist, zhang2023pace, wang2024rcagent, chen2019continuous, hamadanian2023holistic, huang2024faultprofit, shan2024face, wang2024netassistant, xie2024cloud, gao2020scouts, shetty2022autotsg, ML-shetty2021neural, ML-yang2023diffusion,ML-zhao2020real, chen2020towards, he2022graph, pei2025flowofaction}. 
Traditional root cause analysis often employs deep learning models trained on historical incident data for prediction~\cite{ML-shetty2021neural, ML-zhao2020real, ML-yang2023diffusion, chen2019continuous, huang2024faultprofit}. 
Chen et al.~\cite{chen2020towards} provided an early vision for intelligent incident management, and He et al.~\cite{he2022graph} proposed graph-based methods for incident extraction and diagnosis.
More recently, LLMs have been explored for automated incident diagnosis, primarily using retrieval-based methods to match incoming incidents with historical records~\cite{jiang2024xpert, li2023exploring, an2024nissist, zhang2023pace, hamadanian2023holistic, wang2024netassistant, xie2024cloud}. 
Advanced approaches now utilize LLM agents for iterative diagnosis, combining retrieval with feedback-driven analysis via tools for logs~\cite{roy2024exploring, jiang2025l4,zhang2025agentfm}, configuration update~\cite{wang2025identifying}, or physical environment interaction~\cite{wang2024rcagent,hamadanian2023holistic}. Flow-of-Action~\cite{pei2025flowofaction} further enhances multi-agent root cause analysis with SOP-guided workflows.
Systems such as Minder~\cite{deng2025minder}, MegaScale~\cite{Anomaly-jiang2024megascale}, and Aegis~\cite{dong2025evolution} utilize monitoring metrics to detect failover or failslow scenarios in large-scale distributed AI training. XPUTimer~\cite{cui2025xputimer} and FALCON~\cite{wu2024falcon} trace the communication process for detailed diagnosis. Additionally, solutions like Nissist~\cite{an2024nissist}, AutoTSG~\cite{shetty2022autotsg}, and NetAssistant~\cite{wang2024netassistant} automate traditional troubleshooting guides for infrastructure-level incident diagnosis.
However, these efforts still focus on the provider perspective and certain data sources that they utilize are inaccessible to users.
In contrast, \SYS represents a shift from the ``provider-centric'' to ``user-centric'' paradigm, offering two key distinctions: (1) \SYS acts as a \emph{pre-ticket interception layer} at the moment of ticket submission, enabling user-side diagnosis to filter out resolvable issues before submission, whereas existing works~\cite{RCACopilot,pei2025flowofaction} are designed for post-ticket assistance after an incident has been created and assigned; and (2) \SYS employs \emph{active verification} by running verification scripts to confirm hypothesized faults, while systems like RCACopilot~\cite{RCACopilot} primarily rely on one-shot reasoning over static incident descriptions and logs.

\section{Conclusion}
In this paper, we present \SYS, a user-centric system that automates incident diagnosis for AI workloads. 
During the offline phase, \SYS constructs internal knowledge bases from historical on-call experiences.
For online diagnosis, \SYS employs sequential diagnostic pipelines that mimic the diagnostic processes of human experts, providing immediate feedback to users and facilitating the creation of accurate incident tickets for unresolved problems.
Extensive evaluations using real-world incidents demonstrate the effectiveness of \SYS against various baselines.

\begin{acks}
We would like to thank the reviewers for their valuable comments.
This work is supported in part by the Research Grants Council of the University Grants Committee under Grant 2151323 and by The Chinese University of Hong Kong under Grants 4937007, 4937008, 5501329, and 5501517.
\end{acks}

\newpage
\bibliographystyle{ACM-Reference-Format}
\bibliography{main}










\end{document}